\documentclass[useAMS,usenatbib]{mnras}

\usepackage{natbib}
\usepackage{graphicx}
\usepackage{amsfonts,amsmath,amssymb}

\newcommand{\tev}{t_{\mathrm{ev}}}
\newcommand{\mto}{m_{\mathrm{to}}}
\newcommand{\msun}{\ensuremath{M_{\sun{}}}}
\newcommand{\zsun}{\ensuremath{Z_{\sun{}}}}
\newcommand{\nbody}{$N$-body}
\newcommand{\starlab}{\textsc{Starlab}}
\newcommand{\sevn}{\textsc{SEVN}}
\newcommand{\kira}{\textsc{Kira}}
\newcommand{\sapporo}{\textsc{Sapporo}}
\newcommand{\parsec}{\textsc{PARSEC}}
\newcommand{\std}{\ensuremath{\sigma_{\mathrm{3D}}\left(m\right)}}

\newcommand*{\myalign}[2]{\multicolumn{1}{#1}{#2}}

\title[Energy equipartition in open star clusters]{Do open star clusters evolve toward energy equipartition?}

\author[Spera, Mapelli \& Jeffries]{
	Mario Spera,$^{1}$ 
	Michela Mapelli,$^{1,2}$
	Robin D. Jeffries$^{3}$\\
	$^{1}$INAF, Osservatorio Astronomico di Padova, Vicolo dell'Osservatorio 5, I-35122, Padova, Italy\\
        $^2$INFN, Milano Bicocca, Piazza della Scienza 3, I-20126 Milano, Italy\\
	$^{3}$Astrophysics Group, Keele University, Keele, Staffordshire ST5
	5BG, UK\\}

\begin{document}

\maketitle
\bibliographystyle{mnras}

\begin{abstract}
We investigate whether open clusters (OCs) tend to energy equipartition, by means of direct N-body simulations with a broken power-law mass function. We find that the simulated OCs become strongly mass segregated, but the local velocity dispersion does not depend on the stellar mass for most of the mass range: the curve of the velocity dispersion as a function of mass is nearly flat even after several half-mass relaxation times, regardless of the adopted stellar evolution recipes and Galactic tidal field model. This result holds both if we start from virialized King models and if we use clumpy sub-virial initial conditions. 
The velocity dispersion of the most massive stars and stellar remnants tends to be higher than the velocity dispersion of the lighter stars. This trend is particularly evident in simulations without stellar evolution. 
We interpret this result as a consequence of the strong mass segregation, which leads to Spitzer's instability. Stellar winds delay the onset of the instability. Our simulations strongly support the result that OCs do not attain equipartition, for a wide range of initial conditions. 
\end{abstract}
  

\section{Introduction}
According to the equipartition theorem of statistical mechanics \citep{boltzmann1876}, if a system of gas particles is in thermal equilibrium, an equal amount of energy will be associated  (on average) with each independent energy state. By analogy with ideal gases, a stellar system is expected to evolve toward energy equipartition. In this state, the kinetic energy of a star does not depend on its mass, i.e. $m_i\,{}v_i^2 = m_j\,{}v_j^2$, where $i$ and $j$ are two different particles of the system with mass $m_i$, $m_j$ and velocity $v_i$ and $v_j$, respectively; therefore, the velocity dispersion $\sigma_i$ of the $i$-th group of stars, with mean mass $\widetilde{m_i}$, scales as $\widetilde{m_i}^{-0.5}$.

Two-body encounters are the physical mechanism that allows a stellar system to reach equipartition. On average, a star belonging to a system of $N$ objects undergoes $N\left(1-\frac{4}{N^2}\right)$ encounters per crossing time and each of them alters the kinetic energy  of the particle \citep{binney2008}. Thus, we expect that the typical time-scale to reach thermal equilibrium is comparable to the two-body relaxation time.

The close connection with two-body dynamics makes the equipartition theorem a  local law. Indeed, star clusters are centrally concentrated systems, therefore, their relaxation time in the core is shorter than that of outer regions. Thus, we expect that the core attains energy equipartition faster than the outskirts. In other words, the relation  $m_i\,{}v_i^2 = m_j\,{}v_j^2$ holds at every fixed distance from the cluster centre (i.e. {\em locally}), but it cannot be applied {\em globally} to the entire stellar system.

\cite{spitzer1969} showed that not all stellar systems can reach kinetic energy equipartition: if a stellar system is composed of two stellar populations with stellar mass $m_1$ and $m_2$, with $m_2 \gg m_1$ and $M_2 \ll M_1$ (where $M_1$ and $M_2$ are the total mass of population 1 and 2, respectively), equipartition is possible only if 
\begin{equation}\label{eq:spitzer}
M_2 < 0.16 M_1 \left(\frac{m_2}{m_1}\right)^{-\frac{3}{2}}.
\end{equation}
If $M_2$ is larger than this value, massive stars cannot transfer enough kinetic energy to light stars to reach equipartition. Thus, massive stars kinetically decouple from light stars, and form a self-gravitating system `within the system' (i.e. at the centre of the stellar system), where they interact only with each other. In this configuration, massive stars tend to increase their velocity dispersion.

According to \citet{vishniac1978}, if a continuous distribution of stellar masses is accounted for, equipartition can be reached  only if 
\begin{equation}
M_> < \beta_V M \left(\frac{m}{m_l}\right)^{-\frac{3}{2}},
\label{eq:intro:vishnicac}
\end{equation}
where $m$ is the single stellar mass of a generic group of stars, $m_l$ is the lightest star of the system, $M_>$ is the total mass in stars with mass greater than $m$, and $M$ is the total mass of the system. Equation \ref{eq:intro:vishnicac} is similar to Spitzer's criterion, but $\beta_V$ is an integral function of the density profiles of each stellar mass group and its typical value is $\sim 0.5$. This result implies that, assuming an initial mass function (IMF) such that $\xi\left(m\right)\propto m^{\delta}$, equipartition cannot be attained if $\delta \geq -3.5$. Thus, stellar systems with a realistic IMF ($\delta \simeq -2.3$) cannot reach thermal equilibrium. Equation \ref{eq:intro:vishnicac}  assumes similar density profiles between various stellar mass groups, which might be a poor assumption (e.g. \citealt{merritt1981,stoeger1985}). 
For example, if the assumptions of \cite{spitzer1969} and \cite{vishniac1978} are relaxed, it is always possible to construct theoretical models of two-component stellar systems in thermal and dynamical equilibrium in their cores, regardless of the mass ratio between the two stellar populations \citep{merritt1981}.

Interestingly, the most used model to describe a star cluster, the King model \citep{king1966}, is far from energy equipartition if a realistic mass function is considered \citep{miocchi2006}. The multi-mass Michie-King models \citep{gunn1979} were constructed starting from the assumption of  global energy equipartition, imposing $\sigma^2_i \propto{}m_i^{-1}$ for all the mass classes. However, \citet{merritt1981} and \citet{miocchi2006} showed that the hypothesis of equipartition in a Michie-King model is valid only in the limit of isothermal distribution function, i.e. in systems with central dimensionless potential $W_0 \rightarrow \infty$. Consequently, the actual sampling of a Michie-King model is close to equipartition only if the central density is extremely high \citep{gieles2015}. 


From a numerical point of view, various techniques have been used to investigate equipartition, ranging from Fokker-Planck simulations to direct \nbody{} simulations of two-component stellar systems \citep{spitzer1971,wiyanto1989,inagaki1984,khalisi2007}. Still, only few numerical studies deal with a realistic mass function. 
By means of several direct \nbody{} simulations, \cite{trenti2013} show that kinetic energy equipartition is never reached in globular clusters (GCs). For the $\omega$ Centauri GC, they found $\sigma\left(m\right)\propto m^{-0.16}$, which is in agreement with Hubble Space Telescope (HST) observations \citep{anderson2010}. \citet{bianchini2016} confirm the results discussed by \cite{trenti2013}, showing that the \std{} trend for evolved GCs is exponential.

Understanding whether (or not) star clusters reach energy equipartition is not only a cornerstone of stellar dynamics, but has fundamental implications for a plethora of astrophysics processes. Spitzer's instability dramatically affects the retention of massive stellar remnants in a star cluster, because the kinematically decoupled `sub-cluster within the cluster' is likely made up mostly by black holes (BHs), which are more massive than most stars, after $\sim{}100$ Myr. If BHs interact with each other, several BH binaries can form dynamically, and several BHs (single or in binary systems) can be ejected from the star cluster, because of dynamical kicks \citep{sigurdsson1993,breen2013,sippel2013}. Since mergers of BH binaries are among the most important sources of gravitational waves, the dynamical fate of BHs in star clusters strongly affects the predictions of the detection-rate by ground-based gravitational waves detectors \citep{oleary2006,sadowski2008,downing2010,downing2011,ziosi2014,moerscher2015,rodriguez2015,chatterjee2016,oleary2016}. Similarly,  the dynamics of stellar remnants affects the demographics of X-ray binaries \citep{mapelli2011,downing2011,mapelli2013,berghea2013,mapelli2014,goswami2014}. If equipartition is not reached, there might be relevant implications for the current local mass function of star clusters (e.g. \citealt{beccari2015}), for mass segregation, for dynamical friction, for the dynamical evolution of blue straggler stars (e.g. \citealt{mapelli2004,mapelli2006,ciotti2010,ferraro2012,alessandrini2014}), and the formation of intermediate-mass BHs in star clusters (e.g. \citealt{portegieszwart1999,zwart2002,miller2002,giersz2015,arcasedda2016,mapelli2016}). Thus, understanding equipartition is essential for our knowledge of star clusters.

In this paper we investigate kinetic energy equipartition in open clusters (OCs) by means of direct \nbody{} simulations. We performed several runs of stellar systems composed of $N=6,000$ particles varying initial phase-space conditions. In our simulations we included up-to-date stellar evolution recipes \citep{spera2015} and a static background potential that mimics the Milky Way's tidal field \citep{allen1991}. Since the majority of stars are expected to form in star clusters \citep{lada2003}, OCs  are an optimal target to shed light on a plethora of astrophysical processes, such as star formation, stellar evolution and dynamics of stellar systems. Our models will be soon compared with the data of the Gaia mission \citep{perryman2001} and of the Gaia ESO survey (GES, \citealt{gilmore2012,randich2013}). Gaia is expected to measure  astrometric distances and proper motions  of $\sim{}10^9$ stars. The GES, an ongoing public spectroscopic survey at the Very Large Telescope, aims at measuring line-of-sight velocities and chemistry of $\sim{}10^5$ stars with high accuracy. The combination of Gaia and GES data will provide a 6D phase-space map (plus chemistry information) about the physical properties of $\sim 100$ OCs (up to distances of a few kpc from our Sun). This data set will be the ideal test-bed for understanding the dynamical evolution of star clusters. 

This paper is organized as follows. In Section \ref{sec:nbodysims} we describe the main ingredients of our simulations: the employed \nbody{} code, the open cluster fiducial model, the descriptions of the runs and the data analysis process. In Section \ref{sec:results} we present the results of or simulations in terms of both mass segregation and kinematic state. In Sections \ref{sec:discussion} and \ref{sec:summary} we discuss and summarize our main results.

\section{\nbody{} simulations}
\label{sec:nbodysims}
\subsection{\starlab{} and \sevn{}}
In this paper we investigate the dynamical evolution of OCs by means of direct \nbody{} simulations. 
To run our simulations, we use the \starlab{} software environment \citep{portegieszwart2001}. \kira{}, the direct \nbody{} integrator included in \starlab{}, implements a Hermite 4th order integration algorithm \citep{makino1992} and a neighbors--perturbers scheme to ensure an accurate integration of tight binaries and multiple  systems. We set the softening parameter to zero,  and the radius of a particle to the physical stellar radius (two stars are assumed to merge if their distance is smaller than the sum of the radii). We ran \starlab{} on a Graphics Processing Unit (GPU) nVIDIA GTX Titan Black\footnote{This GPU is based on the nVIDIA Kepler architecture, code name GK110-430-B1.}, by means of the \sapporo{} library v. 1.6 \citep{gaburov2009}. Each run takes, approximately, one hour to be completed.

Stellar evolution is implemented in \starlab{} through \sevn{} \citep{spera2015}. \sevn{} is a tool designed to add stellar
evolution  and  supernova (SN) explosion recipes to \nbody{} simulations. It relies upon a set of input tables extracted from stellar evolution tracks. In this way, if the user wants to change the default stellar evolution tables, he can do it without modifying the internal structure of the \nbody{} code or even recompiling it. The default version of \sevn{} includes the \parsec{} stellar evolution tracks \citep{bressan2012,tang2014,chen2015} and implements several prescriptions for SN explosions \citep{fryer2012,ertl2015}. Moreover, \sevn{} assigns natal kicks to neutron stars (NSs) and BHs according to the three dimensional velocity distribution of the pulsars observed in our Galaxy \citep{hobbs2005}. This value is weighted using the fraction of mass that falls back onto the proto-compact object, so that a BH that forms via direct collapse receives no kicks \citep{fryer2012}.

Moreover, we added  a new recipe for static tidal fields to \starlab{}\footnote{We modified the function \texttt{add\_plummer}, implemented in \texttt{dyn\_external.C}, by adding the contribution of the tidal field to stellar accelerations.}. In particular, we adopted the gravitational potential described in \citet{allen1991}, because it is a simple model of the Milky Way potential, including a spherical, central bulge, a disc  and a massive spherical halo. The central bulge is modeled through a Plummer sphere \citep{plummer1911} whose gravitational potential is
\begin{equation}
\phi_b\left(d\right)=-\frac{GM_b}{\sqrt{d^2+b_b^2}}
\end{equation}
where $d$ is the distance from the Galaxy centre, $M_b=1.41\times 10^{10}\msun{}$ is the total mass of the bulge, and $b_b=0.3873$ kpc. The disc component is represented by a Miyamoto-Nagai disc \citep{miyamoto1975}. The gravitational potential is
\begin{equation}
\phi_d\left(R,z\right) = -\frac{GM_d}{\sqrt{R^2+\left(a_d+\sqrt{(z^2+b_d^2)}\right)^2}} 
\end{equation}
where $R$ is the distance from the Galaxy centre on the x-y plane, $M_d=8.56\times 10^{10}\msun{}$ is the total mass of the disc, $a_d=5.3178$ kpc and $b_d = 0.25$ kpc. The Galaxy halo is modeled through a spherical logarithmic potential of the form
\begin{equation}
\begin{split}
\phi_h\left(d\right)=&-\frac{GM_h\alpha^{2.02}}{d\left(1+\alpha^{1.02}\right)}+\\
&-\frac{GM_h}{1.02a_h}\left[-\frac{1.02}{1+\alpha^{1.02}}+\ln{\left(1+\alpha^{1.02}\right)}\right]_d^{100\,{}{\rm kpc}}
\end{split}
\end{equation}
where $M_h=8.002\times 10^{11}\msun{}$ is the total mass of the halo, $a_h=12$ kpc and $\alpha \equiv \dfrac{d}{a_h}$.

\subsection{OC models}

Our fiducial OC \nbody{} model is composed of $N=6000$ stars, whose masses are distributed according to a broken power-law IMF \citep{kroupa2001} with lower mass limit $m_{\mathrm{low}}=0.1$\msun{} and upper mass limit $m_{\mathrm{up}}=150$\msun{}. The slope of the IMF is $\alpha_{\mathrm{1}}=1.3$, for $m_{\mathrm{low}} \leq m < 0.5$\msun{}, and $\alpha_{\mathrm{2}}=2.3$, for $0.5$\msun{}$\leq m \leq m_{\mathrm{up}} $. As a consequence, the average initial mass of our cluster is $M\simeq 3900$\msun{}. We assign  a slightly super-solar metallicity\footnote{We consider the value $\zsun{}=0.01524$ as solar metallicity, according to \citet{caffau2011}.}, $Z=0.02$, to our fiducial OC model. We do not include primordial binaries, but tight binaries and multiple systems can form during the numerical integration and they are  handled by the neighbors-perturbers  module of the \kira{} integrator.

The initial positions and velocities are sampled from a \cite{king1966} distribution function with central dimensionless potential $W_0=5$ and King's core radius $r_0=0.4$ pc, which corresponds to a concentration $\simeq{}1.03$ and an initial half-mass radius $r_h\simeq0.8$ pc. The initial half-mass relaxation time $t_{\rm rh}\left(0\right)$ is \citep{spitzer1969}
\begin{equation}
\label{eq:hm_reltime}
t_{\rm rh}\left(0\right)=\frac{0.17N}{\mathrm{ln}\left(\lambda N\right)}\sqrt{\frac{r_h^3}{GM}}.
\end{equation}
\cite{giersz1994} suggested $\lambda \simeq 0.1$. For our \nbody{} model, this formula gives $t_{\rm rh}\left(0\right) \simeq 27$ Myr. The core collapse time of a stellar system whose stars are distributed according to a realistic mass spectrum is
\begin{equation}
\label{eq:corecollapse}
t_{\rm cc}=\gamma t_{\rm rh}
\end{equation}
where $\gamma = 0.1 \div 0.2$ \citep{zwart2002,fujii2014}. We expect core collapse for our fiducial OC at time $3\lesssim t_{\rm cc}/\mathrm{Myr}\lesssim 5$.

All  simulations stop at $t_{\mathrm{ev}}= 160$ Myr, which means that we evolved the stellar system for $\sim 6$ $t_{\rm rh}\left(0\right)$ (in Appendix~A, we discuss what happens at later times, up to $\tev{}= 1$ Gyr). 

We expect that the time-scale for kinetic energy equipartition scales as the mass-segregation timescale, which, in turn, is connected with the two-body relaxation time-scale. The mass segregation time-scale  for a star of mass $\widetilde{m}$, inside a star cluster  composed of stars with average mass $\left<m\right>$, is given by \cite{spitzer1969}
\begin{equation}
\label{eq:seg_time_m}
t_{\mathrm{seg}}\left(\widetilde{m}\right) \simeq \frac{\left<m\right>}{\widetilde{m}}t_{\rm rh}\left(0\right).
\end{equation}
Thus, we expect that stars with mass 
\begin{equation}\label{eq:msegr}
M_{\mathrm{seg}}\ge{}0.1\,{}\msun{}\,{}\left(\frac{\langle{}m\rangle{}}{0.6\,{}\msun{}}\right)\,{}\left(\frac{160\,{}{\rm Myr}}{t_{\mathrm{ev}}}\right)\,{}\left(\frac{t_{\rm rh}(0)}{27\,{}{\rm Myr}}\right) 
\end{equation}
have reached equipartition by the end of the simulation. Since $M_{\mathrm{seg}}\simeq m_{\mathrm{low}}$ (i.e. the minimum stellar mass adopted in our simulations), the simulated \nbody{} systems are expected to attain mass segregation and energy equipartition by the end of the simulation, at least inside the half-mass radius, and for a wide range of masses. 

Actually, the true minimum mass of stars that segregated to the centre might be slightly larger than the value of 
$M_{\mathrm{seg}}$, because equation~\ref{eq:msegr} neglects the time evolution of $t_{\rm rh}$, due to the change of the total mass $M$ and half mass radius $r_{h}$. For example, for our fiducial runs $t_{\rm rh}(160\,{}{\rm Myr})\sim{} 215$ Myr. Hence the upper limit of $M_{\mathrm{seg}}$ is $\sim 0.6\msun{}$, still close to the minimum mass of stars in our simulations.

\subsection{Description of runs}
\label{subsec:desc_runs}
  In this paper, we present the results we obtained from the following four groups of simulations.
\begin{itemize}
\item Group A: we include both stellar evolution (as described in \citealt{spera2015}) and the effect of the Galactic tidal field (using the potential  described in \citealt{allen1991}). 
\item Group B: we include stellar evolution, while the tidal field contribution is switched off: the star cluster evolves in isolation.
\item Group C: both stellar evolution and the tidal field are switched off.
\item Group D: as in group~A, stellar evolution and tidal field contributions are included, but we start from a completely different OC model.
Instead of simulating a monolithic King model, the initial conditions for each individual star cluster are obtained by generating 20 sub-clusters, each composed of 300 particles. Each sub-cluster is sampled from a \cite{king1966} distribution function with $W_0=2$ and $r_0=0.2$ pc. The centres of mass of the sub-clusters are distributed homogeneously in a sphere of radius 10 pc and  have null initial velocity. The resulting star cluster has the same number of particles ($N=6000$) as in the monolithic star cluster models, but, initially, the system is not in virial equilibrium. The aim of the simulations of group~D is to try to mimic realistic initial conditions for young OCs, as recent observations suggest that young stellar systems are subvirial and clumpy aggregations of several sub-clusters \citep{mcmillan2007,schmeja2008,proszkow2009,spera2015b}. Moreover, the simulations of Group D allow us to check whether our results depend on the initial conditions.
\end{itemize} 
 
Table \ref{tab:tab1} summarizes the properties of our simulations. 
Each group of simulations is composed of $200$ realizations of the same star cluster, to filter out statistical fluctuations.

\begin{table} 
\begin{center}
\caption{\label{tab:tab1} Properties of the four groups of simulations.}
    \begin{tabular}{ c c c c c }
    \hline
         & Tidal field & St. evo. & Clumpy IC & King $W_0$ \\ 
    \hline
        A & $\surd$ & $\surd$ & $\times$ & 5 \\ 
        B & $\times$ & $\surd$ & $\times$ & 5  \\ 
        C & $\times$ & $\times$ & $\times$ & 5  \\ 
        D & $\surd$ & $\surd$ & $\surd$ & 2$^\mathrm{a}$\\
    \hline
    \end{tabular} 
\end{center}
\footnotesize{$^\mathrm{a}$ $W_0$ parameter of each clump.}
\end{table}

The orbital parameters of the star cluster in the Galactic tidal field have been chosen to match the orbit of the nearby, intermediate-age OC NGC 2516 \citep{wu2009}\footnote{The complete catalog can be found at \url{http://vizier.u-strasbg.fr/viz-bin/VizieR?-source=J/MNRAS/399/2146}}, which is one of the targets of the GES. In particular, in the simulations of groups A and D, we placed the centre of mass of the \nbody{} system at position $\mathbf{r}_{\mathrm{cl}} = \left(-7.974; -0.393; -0.112\right)$ kpc with velocity $\mathbf{v}_{\mathrm{cl}}=\left(-8.5; 200.7; 4.3\right)$ km/s with respect to the centre of the Galactic potential. Moreover, these simulations are designed to be comparable with a number of rich young open clusters (e.g. the Pleiades, NGC 2516, M35) at ages of $100-200$ Myr for which data from GES and Gaia will shortly be available (Appendix~A shows what happens for older OCs, up to $\tev{}= 1$ Gyr, in runs of group~A).

 The initial filling factor of the star cluster  for the chosen orbital set-up is $r_{\rm h}/r_{\rm J} \simeq 0.028$ where $r_{\rm J}$ is the Jacobi radius (e.g. \citealt{king1962}) defined as
	\begin{equation}
	r_{\rm J} = \left(\frac{G\,{}M}{V_G^2}\right)^{\frac{1}{3}}r_{\mathrm{cl}}^{\frac{2}{3}}
	\end{equation}
where $V_G$ is the circular velocity of the galaxy at distance $r_{\mathrm{cl}}$.


	\begin{table} 
		\begin{center}
			\caption{\label{tab:tab1a} Median mass bound to an OC ($M_{\rm bound}$), corresponding absolute error ($s\left(M_{\rm bound}\right)$), and half-mass radius $r_{\mathrm{h}}$, at $t= 160$ Myr, for all the simulation groups.}
			\begin{tabular}{ c c c c }
				\hline
				Group & $M_{\rm bound}$ & $s\left(M_{\rm bound}\right)$ & $r_{\mathrm{h}}$ \\
				 & ($10^3$ M$_\odot$) & ($10^3$ M$_\odot$) & pc\\
				\hline
				A & 2.675 & 0.033  & 3.2\\ 
				B & 2.67 & 0.10 & 3.2\\ 
				C & 2.62 & 0.18 & 8.3\\ 
				D & 2.305 & 0.063 & 4.6 \\
				\hline
			\end{tabular} 
		\end{center}
	\end{table}
	
	Tab.~\ref{tab:tab1a} shows the median bound mass $M_{\rm bound}$ of the simulated OCs at $t_{\rm ev}=160$ Myr. In the OCs of groups A and D, a star is considered unbound  if (i) the distance between the star and the OC centre is more than $2.0\,{}r_{\rm J}$, and (ii) the star is moving away from the OC. In the OCs of group~B and C (without tidal field), we adopt the same criteria but we use  the cluster half-mass radius  instead of $r_{\rm J}$. Tab.~\ref{tab:tab1a} shows that OCs of groups A, B, and C have approximately the same final mass ($\sim 2600\msun{}$), which means that they have lost $\sim 1/3$ of their initial mass. The OCs of group~D lose more mass as a consequence of the initial violent collapse of the 20 sub-clusters.

\subsection{Data analysis}
\label{sec:dataanalysis}
We performed $200$ realizations of the same initial conditions and each run generates,  approximately, 160 snapshots, corresponding to one output every $\sim 1$ Myr. 
To quantify whether a stellar system is in thermal equilibrium, we evaluate the three-dimensional velocity dispersion of stars as a function of their mass \std{}. We can get this information from our simulations following two alternative approaches: the first approach is based on the median, while the second one is based on the stack technique. 

In the {\em median} approach, we first evaluate \std{} for each snapshot of each run, dividing stars into several mass groups. Then, at fixed time and mass bin, we collect all the \std{} values from runs in the same group. The final estimation of \std{}, at that specific time and mass bin, is the median  of all the collected values and the associated error is their standard deviation.

In the {\em stack} approach, at each time we merge all the snapshots from all runs belonging to the same group in a single file. Then, we evaluate \std{} from that file, dividing stars into several mass groups. To estimate the error, we have considered the Poissonian uncertainty associated with the number of particles in each mass bin, per star cluster, and we propagate it following the standard propagation formula.

 We checked that the results of the two methods are consistent. To quantify their confidence level we perform a z-test (e.g. \citealt{frederick2006}). In particular, we evaluate the maximum value of the variable
\begin{equation}
	\mathcal{Z} = \frac{\left|x_m-x_s\right|}{\sqrt{s_m^2+s_s^2}}
\end{equation}
where $x_m$ and $x_s$ are two generic measures of \std{} obtained using the {\em median} and the {\em stack} approach for the same mass bin, respectively, and $s_m$ and $s_s$ are the corresponding errors. For the simulations of group~A we obtain
\begin{equation}
\max{\mathcal{Z}} = 0.39.
\end{equation}
Thus, the probability $P\left(\mathcal{Z}\right)$ of observing a standard normal value $>0.39$ and $<-0.39$ is $P\left(\mathcal{Z}\right) = 70\%$. This implies that the {\em median} and the {\em stack} approaches show high compatibility. In this paper, we choose to present the results we obtained using the {\em stack} method because it allows us to slightly reduce statistical fluctuations.

We estimate \std{} in $10$ mass bins (unless otherwise specified), which are logarithmically distributed  between $m_1=0.1\msun{}$ and $m_2=25\msun{}$. In particular, $m_1 = 0.1\msun{}$ is the lower limit of the initial mass function of our stellar systems, while $m_2=25\msun{}$  is, roughly, the mass of heaviest BH formed in our simulations (e.g. \citealt{spera2015}).
To reduce the influence of statistical fluctuations, we exclude the mass bins that contain less than 3 particles per star cluster.

Unless otherwise specified, \std{} was derived substituting binary systems with their centres of mass. This is important because if we plot the radial velocity of each binary member (including the radial component of the orbital velocity), we can get a spurious overestimate of the local velocity dispersion. Identifying binaries and accounting for the orbital velocity is one of the most serious problems when measuring the velocity dispersion from observational data.

Finally, to avoid spurious effects in the determination of the physical parameters of the simulated stellar systems, we only consider the particles inside the Lagrangian radius containing 50\% of the total system mass, unless otherwise specified.

Moreover, we want to quantify the degree of mass segregation in the simulated OCs. To do this, we use the minimum spanning tree (MST) method. We recall that the MST is the shortest path length which connects a certain number of points without forming closed loops. The MST technique is one of the most useful methods to quantify mass segregation in stellar systems and does not depend either on the geometry or on the position of the centre of the star cluster \citep{allison2009,parker2015}. We use the notation $\mathcal{M}_{>m}$  and $N_{>m}$ (subscripts are in units of \msun{}) to indicate the MST length and the number of stars with mass larger than $m$, respectively, while $\widetilde{\mathcal{M}_{N}}$ indicates the average MST for a sample of $N_{>m}$ randomly selected stars in the whole system. We use the parameter $\Lambda_{>m}\equiv \left(\frac{\widetilde{\mathcal{M}_{>m}}}{\mathcal{M}_{>m}}\right)$ to quantify the degree of mass segregation for stars with mass larger than $m$. Values $\Lambda_{>m}>1$ indicate that the population of stars with mass larger than $m$ is more concentrated than the average. In particular, if the ratio $\frac{\Lambda_{>m_1}}{\Lambda_{>m_2}} > 1$ then the population number 1 is more segregated than population number 2.

As complementary information to MSTs, we also evaluated the radial profiles of different classes of mass. We indicate with $\mathcal{N}_{>m}\left(r\right)$ the number of stars with mass larger than $m$, at a distance $r$ from the centre of density of the stellar system. The quantity we used to infer information on mass distribution is the following normalized curve:
\begin{equation}
\mathcal{K}_m\left(r\right) \equiv \frac{\mathcal{N}_{>m}\left(r\right)}{\mathcal{N}_{<0.2}\left(r\right)}\frac{N_{<0.2}}{N_{>m}}.
\label{eq:massdistribution}
\end{equation}
$\mathcal{K}_{m_1}\left(\widetilde{r}\right) > \mathcal{K}_{m_2}\left(\widetilde{r}\right)$ means that, at distance $r=\widetilde{r}$, the population of stars with mass $m_1$ has larger relative abundance than population with mass $m_2$.

\section{Results}
\label{sec:results}

\subsection{Mass segregation}
\begin{figure}
	\begin{center}
		\includegraphics[width=\columnwidth]{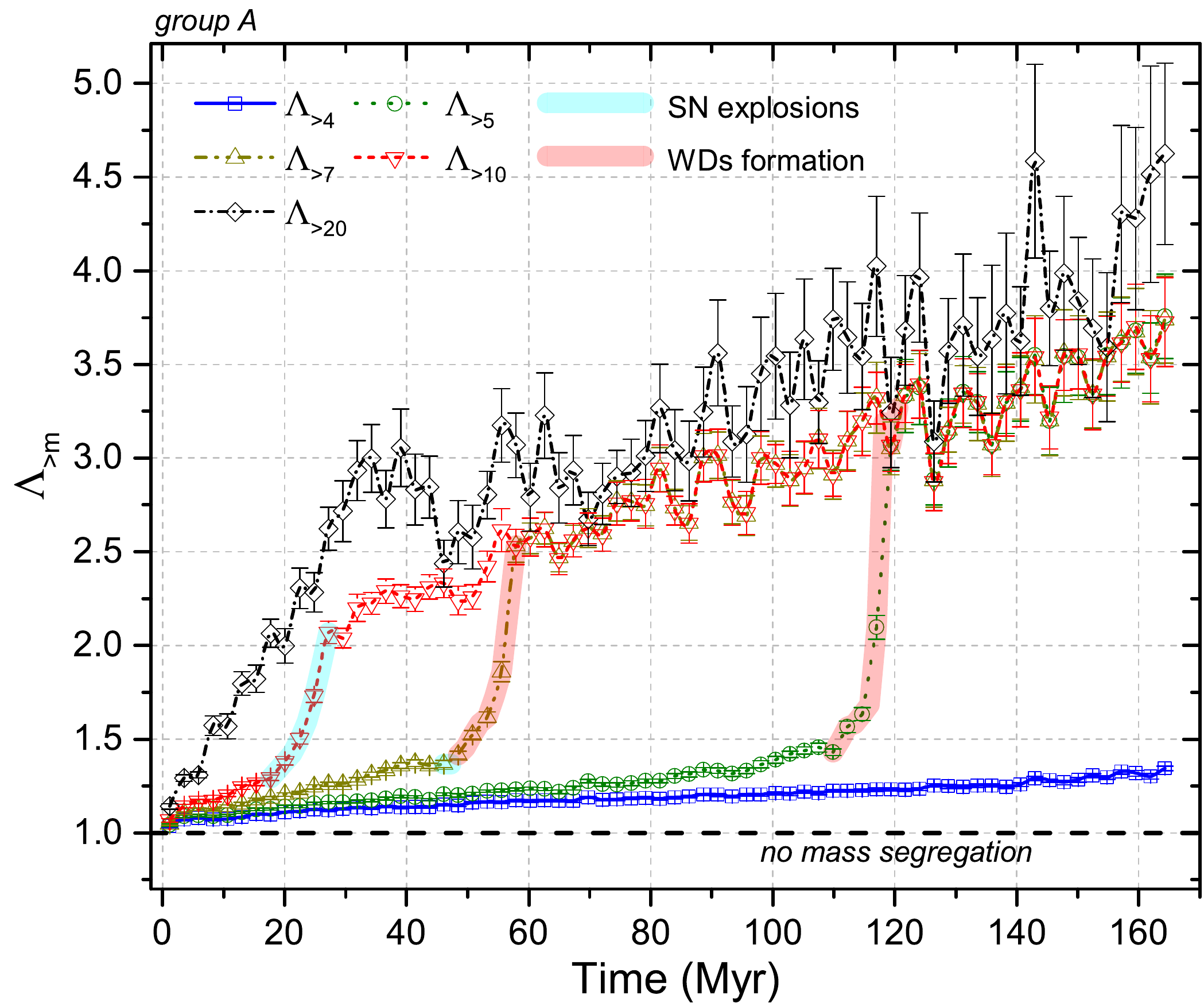}
		\caption{\label{fig:fig1} $\Lambda_{>m}$ parameter, as a function of time (in Myr), for the simulations of group~A. The dashed black line at $\Lambda_{>m}=1$ indicates the absence of mass segregation. Solid blue line (with squares): stars with mass $m>4\msun{}$. Dotted green line (with circles): stars with $m>5\msun{}$. Dash-dot-dot dark yellow line (with upward triangles): stars with $m>7\msun{}$. Dashed red line (with downward triangles): stars with $m>10\msun{}$. Dash-dot black line (with rhombi): stars with $m>20\msun{}$. The thick, semi-transparent cyan and red lines highlight the intervals in which SN explosions and the formation of WDs, respectively, occur. Here, $m$ does not refer to the initial mass of a star, but to its mass at the time shown on the $x-$ axis.}
	\end{center}
\end{figure}

Fig. \ref{fig:fig1} shows the value of the parameter $\Lambda_{>m}$ (see Sec. \ref{sec:dataanalysis}) as a function of time, for the simulations of group~A, for stars with masses $m>4\msun{}$, $m>5\msun{}$, $m>7\msun{}$, $m>10\msun{}$ and $m>20\msun{}$. Here, $m$ does not refer to the initial mass of a star, but to its mass at the time shown on the $x-$ axis. At the beginning of the simulation, $\Lambda_{>m}\simeq 1$ for all the considered mass groups. This means that, at $t=0$, the stellar system is not mass segregated.  

At the end of the simulation, $\Lambda_{>m}> 1$ for all the considered mass groups: the stellar system is clearly mass segregated. The $\Lambda_{>m}$ parameter grows with time for all the classes of mass and, at the end of the simulation, reaches a value $\simeq 4.5$ for stars with $m>20\msun{}$, while it is only $\sim 1.3$ for stars with $m>4\msun{}$. As expected, the process of mass segregation is particularly efficient for massive stars. The curves $\Lambda_{>10}$,  $\Lambda_{>7}$ and $\Lambda_{>5}$ show an abrupt step which starts at times 17 Myr, 45 Myr and 95 Myr, respectively. This rapid variation corresponds to the beginning of the SN explosions (for stars with mass $m\gtrsim 8 \msun{}$) and to the beginning of the formation of carbon-oxygen white dwarfs (WDs, for $m \lesssim 8\msun{}$). According to the stellar evolution recipes and SN explosion prescriptions used in our simulations \citep{spera2015,bressan2012,fryer2012}, stars with mass $5\lesssim m/\msun{} \lesssim 20$ form compact remnants with masses $m_{\mathrm{cr}} \lesssim 3 \msun{}$. Thus, after the formation of compact remnants, the curves $\Lambda_{>10}$,  $\Lambda_{>7}$ and $\Lambda_{>5}$ tend to resemble the curve $\Lambda_{>20}$ since the majority of stars with mass $5\lesssim m/\msun{} \lesssim 20$ have become light compact objects (NSs or WDs) with mass $m_{\mathrm{cr}} \lesssim 3 \msun{}$. We do not see the same abrupt step in the curve $\Lambda_{>4}$ since stars with $m\lesssim 4.8 \msun{}$ are still in the main sequence at the end of our simulations.

\begin{figure}
	\begin{center}
		\includegraphics[width=\columnwidth]{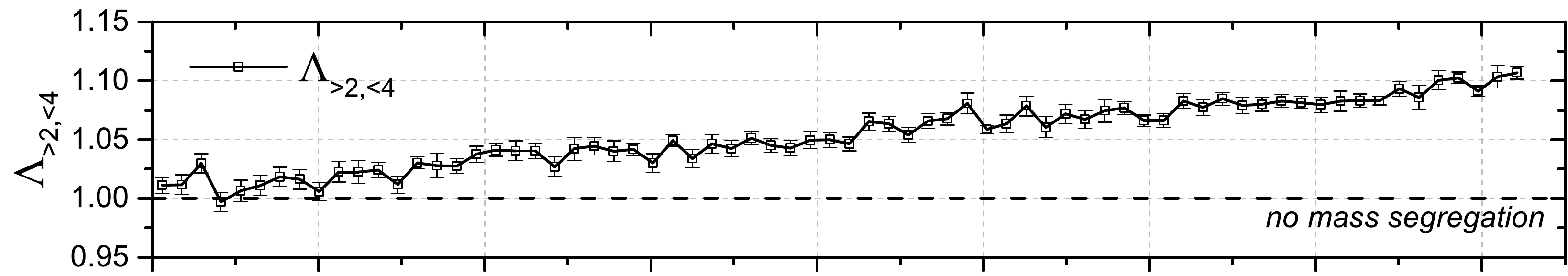}
		\includegraphics[width=\columnwidth]{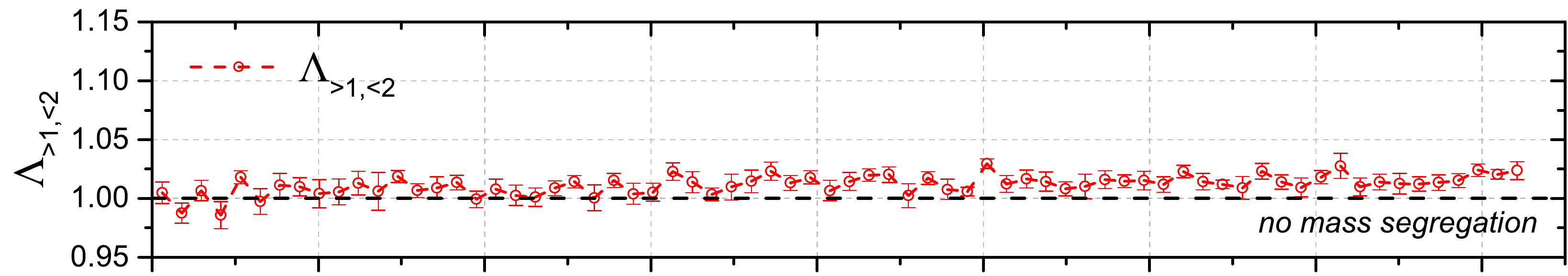}
		\includegraphics[width=\columnwidth]{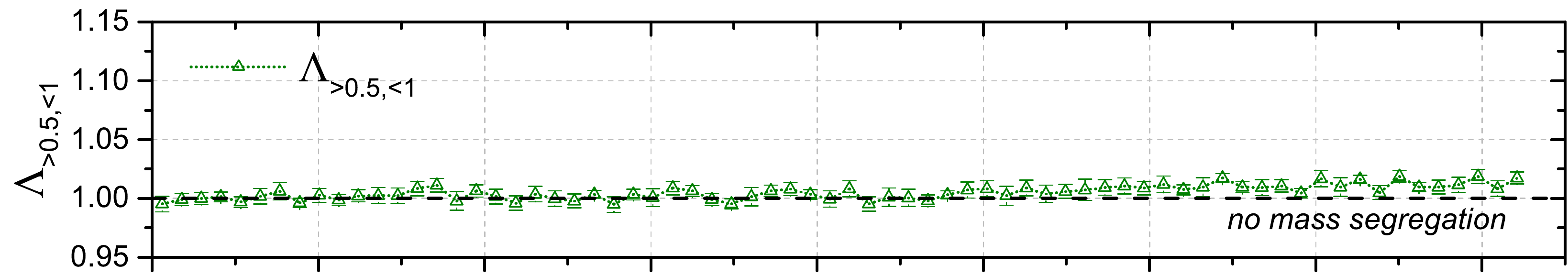}
		\includegraphics[width=\columnwidth]{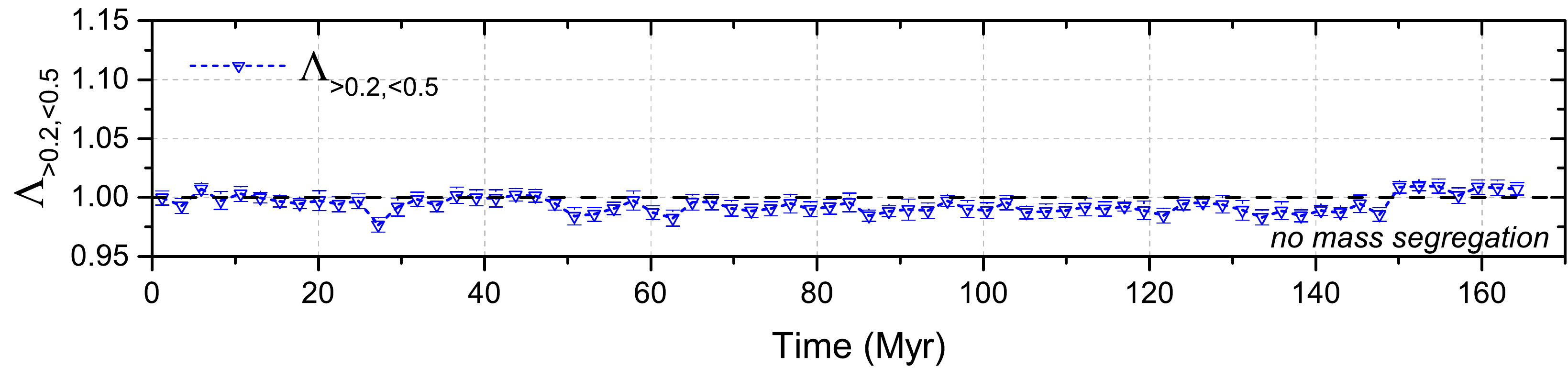}
		\caption{\label{fig:fig2} $\Lambda_{>m_1,<m_2}$ parameter, as a function of time (in Myr), for the simulations of group~A, for different mass ranges. First panel from top: stars with mass $2\leq m/\msun{}\leq 4$; second panel: $1\leq m/\msun{}\leq 2$; third panel: $0.5\leq m/\msun{}\leq 1$; forth panel: $0.2\leq m/\msun{}\leq 0.5$. The dashed black line at $\Lambda_{>m_1,<m_2} = 1$ indicates the absence of mass segregation.}
	\end{center}
\end{figure}

Fig. \ref{fig:fig2} shows the MST curves, as a function of time, for stars in several mass ranges between $0.2$ and $4\,{}\msun{}$. 
	At $t\simeq 160$ Myr, stars with mass $2\leq m/\msun{}\leq 4$ are mass-segregated, whereas stars with mass $1\leq m/\msun{}\leq 2$ are only marginally mass-segregated. Stars with masses $0.5\leq m/\msun{}\leq 1.0$ and $0.2\leq m/\msun{}\leq 0.5$ are not mass segregated.  Thus, the OCs of group~A are mass segregated down to $\sim 1\div 2\msun{}$ at the end of the simulation.

\begin{figure}
	\begin{center}
		\includegraphics[width=\columnwidth]{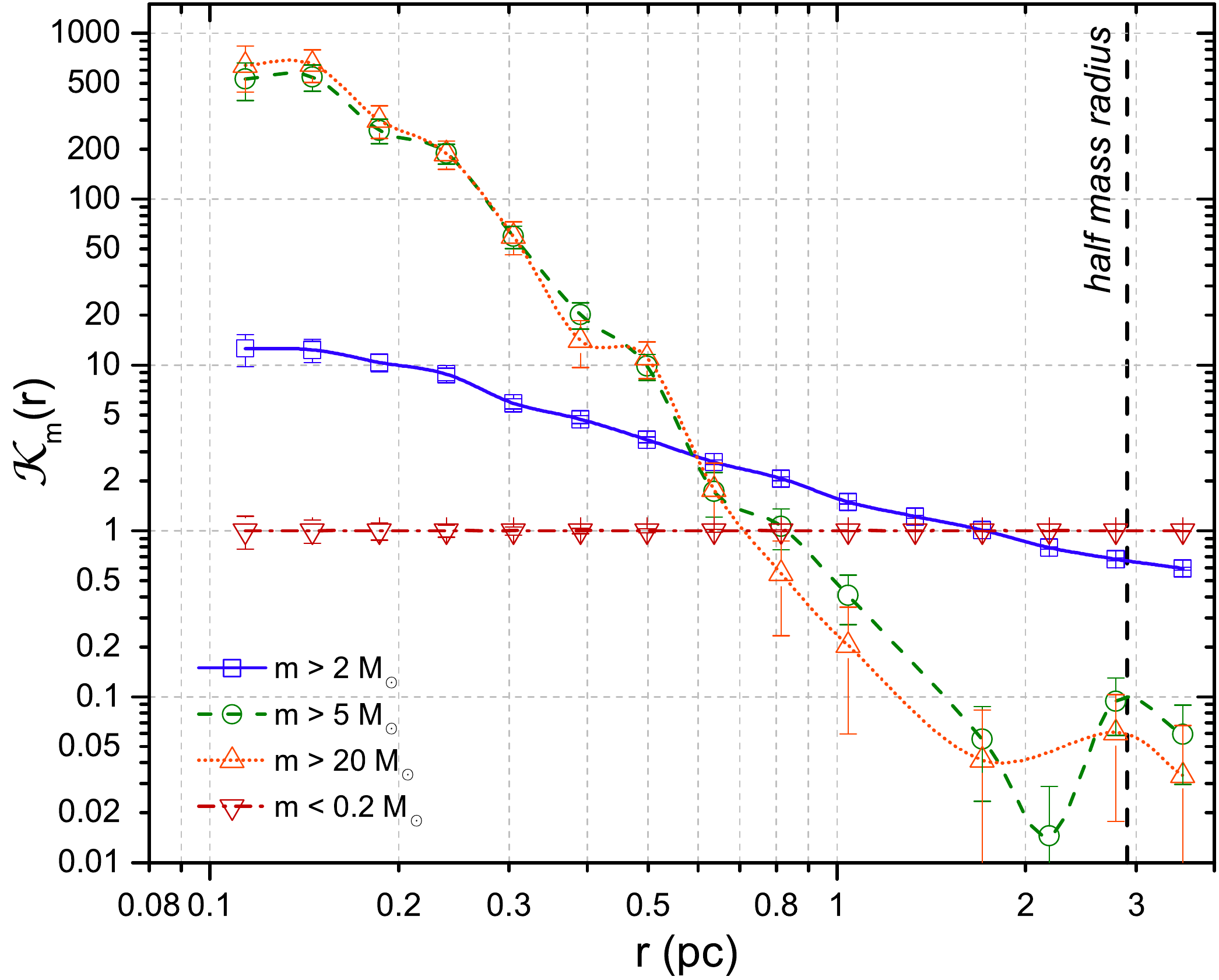}
		\includegraphics[width=\columnwidth]{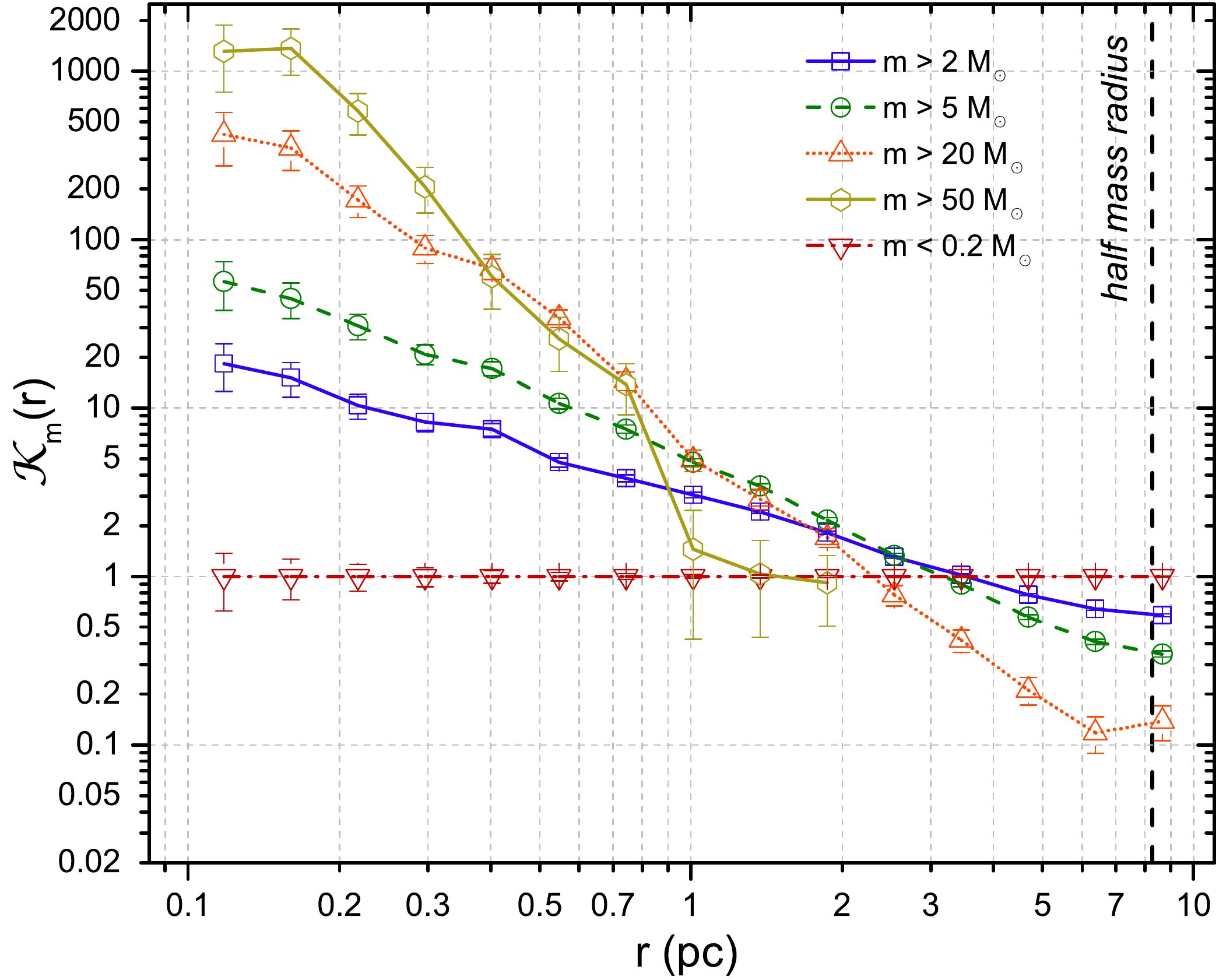}
		\caption{\label{fig:fig3} $\mathcal{K}_m$ parameter as a function of the distance from the centre of density of the stellar system (in pc), for the simulations of group~A (top panel) and  C (bottom panel), at the end of the simulation (160 Myr). The dashed black vertical line at $r\simeq 3.5$ pc indicates the half mass radius of the stellar system, at $t=160$ Myr. Solid blue line (with squares): stars with mass $m>2\msun{}$. Dashed green line (with circles): stars with $m>5\msun{}$. Dotted orange line (with upward triangles): stars with $m>20\msun{}$. Solid ochre line (with hexagons): stars with $m>50$ M$_\odot$ (only in group~B, bottom panel). Dash-dot red line (with downward triangles): stars with $m<0.2\msun{}$.}
	\end{center}
\end{figure}

As a complementary information to spanning trees,  Fig. \ref{fig:fig3} shows the mass distribution  $\mathcal{K}_m\left(r\right)$, derived from equation~\ref{eq:massdistribution},  as a function of the distance from the centre of density of the OC. The top panel of Fig. \ref{fig:fig3} refers to the simulations of group~A, at the end of the runs (160 Myr). The bottom panel refers to simulations of group~C, which are the ones that differ more from group~A, because they do not include stellar evolution. Simulations of group~B and D are not shown in this figure, because mass segregation proceeds in the same way as in simulations of group~A.

In both group~A and group~C, the heaviest stars are much more abundant in the inner region of the stellar system ($r\lesssim 0.5$ pc), while the stellar population with $m<0.2 \msun{}$ becomes more abundant in the outer regions ($r\gtrsim 2 $ pc). This confirms that the stellar system is mass segregated at the end of the simulation. The evidence of mass segregation in our simulations confirms that the process of dynamical friction is efficient, as expected from analytic calculations (see equation~\ref{eq:seg_time_m}). In the simulations of group~C, stars with $m\sim{}5-20$ M$_\odot$ are less mass segregated than in simulations of group~A, because simulations of group~C contain stars that are much heavier than $\sim{}20$ M$_\odot$. In group~C, stars with mass $m>50$ M$_\odot$ are significantly more segregated than the other classes of mass.

\subsection{Kinematic state}

\begin{figure}
	\begin{center}
		\includegraphics[width=\columnwidth]{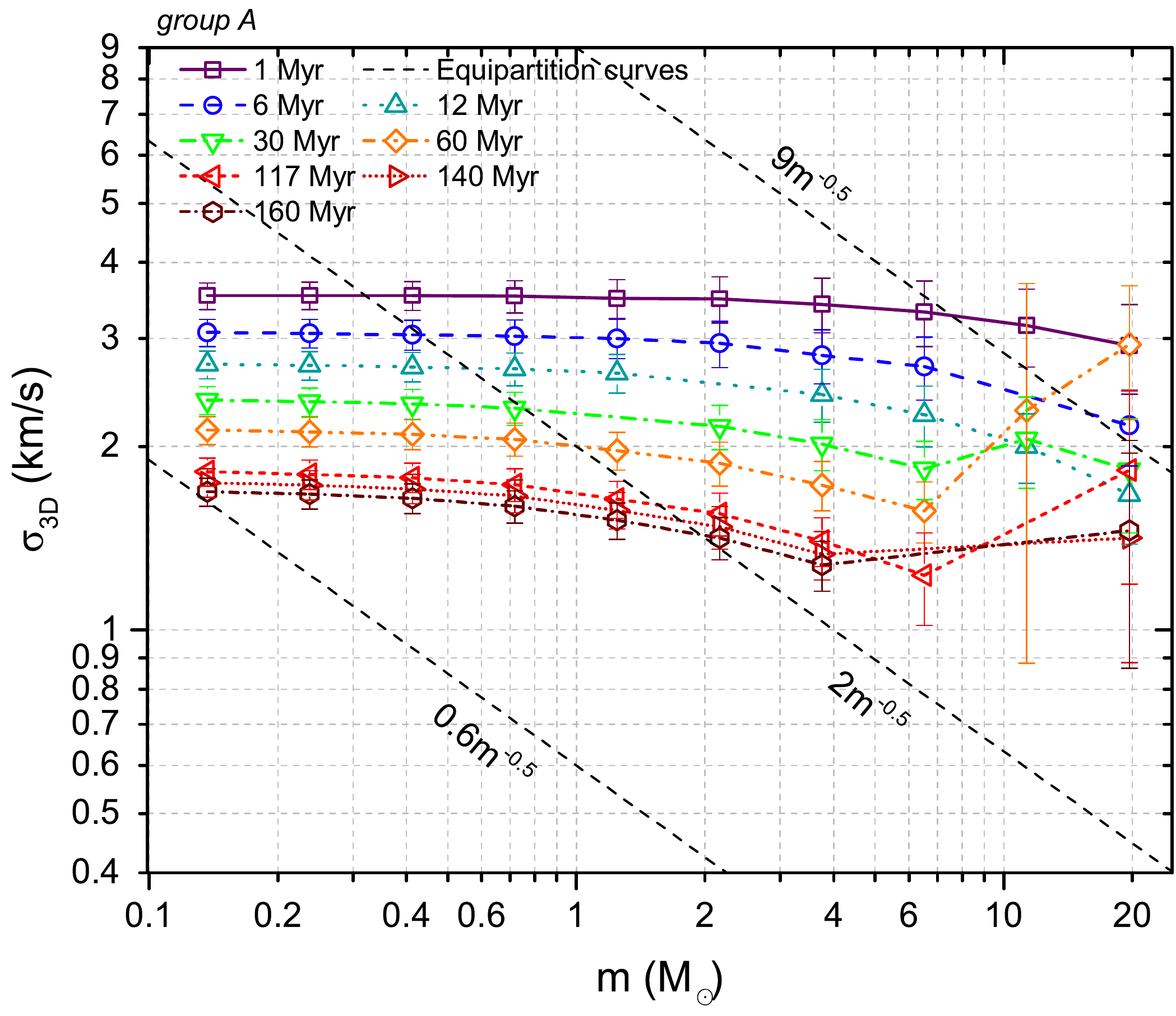}
		\caption{\label{fig:fig4}  Three-dimensional velocity dispersion \std{}, as a function of mass, for the simulations of group~A, at different selected times.  Purple solid line (with squares): time $t=1$ Myr; dashed blue line (with circles): $t=6$ Myr;  teal dotted line (with upward triangles):  $t=12$ Myr; dash-dot green line (with downward triangles): $t=30$ Myr; dash-dot-dot orange line (with rhombi): $t=60$ Myr; short-dashed red line (with leftward triangles): $t=117$ Myr; short-dotted dark red line (with rightward triangles): $t=140$ Myr; dash-dot brown line (with hexagons): $t=160$ Myr. The dashed black lines are the family of equipartition curves $\sigma\left(\alpha;m\right) = \alpha m^{-0.5}$.}
	\end{center}
\end{figure}

Fig. \ref{fig:fig4} shows the three-dimensional velocity dispersion as a function of mass \std{}, at different selected times, for the simulations of group~A. It is apparent that the stellar system is far from energy equipartition during the entire simulation.  

Initially, the \std{} curve is approximately flat  since the velocity distribution of stars does not depend on mass, by construction. As the system evolves with time, stars with mass $m \geq 10\msun{}$ try to reach energy equipartition and, at $t \simeq 12$ Myr, their velocity dispersion is  consistent with thermal equilibrium. However, at $t \simeq 30$ Myr, the \std{} of massive stars seems to rise up and breaks their equipartition state. At $t \simeq 60$ Myr, the \std{} curve of heavy stars has increased its value while there are no significant changes in the \std{} trend of light stars. Stars with masses in the range $4\msun{} \lesssim m \lesssim 6\msun{}$ are the closest ones to equipartition. At $t \simeq 117$ Myr  the velocity dispersion of massive stars decreases again, even if this is within the statistical uncertainties, and it keeps decreasing till $t \simeq 140$ Myr when the \std{} curve seems to reach a stationary state. The final kinematic state is far from equipartition in the entire range of stellar masses, with \std{} $\propto{}m^{-0.07}$ for stars with mass $<5$ M$_\odot$ (see Table~\ref{tab:tab2}). The stars in the mass range $2\msun{} \lesssim m \lesssim 4\msun{}$ are the closest ones to energy equipartition at $t=160$ Myr. The turn-off mass at $t=160$ Myr is $\sim{}5$ M$_\odot$. Thus, stars with masses just below the turn-off tend to be  slightly closer to energy equipartition than the other stars.

\begin{figure}
	\begin{center}
		\includegraphics[width=\columnwidth]{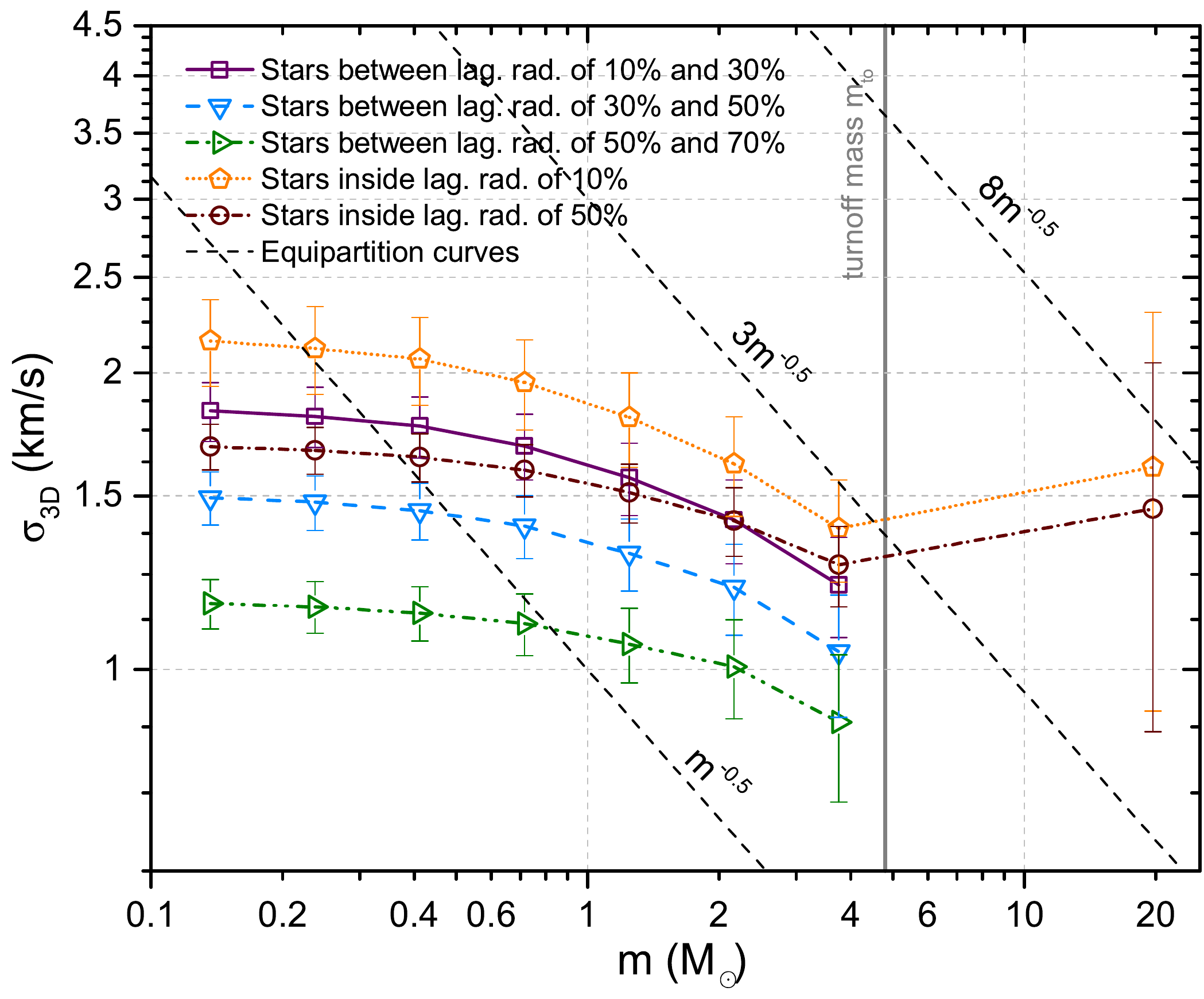}
		\caption{\label{fig:fig5} Three-dimensional velocity dispersion \std{}, as a function of mass, for the simulations of group~A, at time $t=160$ Myr $\simeq 6 t_{\mathrm{rh}}\left(0\right)$, for different areas inside the stellar system. The dashed black lines are the family of equipartition curves $\sigma\left(\alpha;m\right) = \alpha m^{-0.5}$ and the solid grey line  at $m = m_{\mathrm{to}} = 4.8\msun{}$ indicates the turn-off mass at $t = 160$ Myr. Solid purple line (with squares): stars whose distance from the OC centre is $R_{10\%}\le{}r<R_{30\%}$ (where $R_{10\%}$ and $R_{30\%}$ are the radii that enclose 10\% and 30\% of the total OC mass, respectively); dashed blue line (with downward triangles):  stars whose distance from the OC centre is $R_{30\%}\le{}r<R_{50\%}$ (where $R_{50\%}$ is the radius that encloses 50\% of the total OC mass); dot-dashed green line (with rightward triangles):  stars whose distance from the OC centre is $R_{50\%}\le{}r<R_{70\%}$ (where $R_{70\%}$ is the radius that encloses 70\% of the total OC mass); short-dotted orange line (with pentagons):  stars whose distance from the OC centre is $r<R_{10\%}$; dash-dot brown line (with circles): stars whose distance from the OC centre is $r<R_{50\%}$. The latter curve is the same as the dash-dot brown line (with hexagons) shown in Fig. \ref{fig:fig4}.}
	\end{center}
\end{figure}

Fig. \ref{fig:fig5} shows the \std{} trend for the simulations of group~A, at $t=160$ Myr,  at various distances from the centre of the stellar system (defined by various Lagrangian radii). In this way, we can check whether the stellar system attains energy equipartition locally (i.e. in a radial annulus of the OC). This is an important point, since the equipartition principle is valid locally rather than globally. As expected, the average absolute value of the velocity dispersion in the inner regions is higher than that in the outskirts. Still, we find that the stellar system does not show significant differences among different regions in terms of thermal equilibrium. This implies that energy equipartition is not attained either globally or locally. For $m\lesssim 5\msun{}$, we obtain $\std{} \propto m^{-0.1}$ considering the stars inside the core only (Lagrangian radius of 10\%) and $\std{} \propto m^{-0.06}$ for stars in the outer area (between the half-mass radius and the Lagrangian radius of 70\%).  There are no massive objects ($m\gtrsim 5\msun{}$) in the outskirts of the stellar system (outside the half-mass radius, approximately). This confirms that the simulated star clusters are mass segregated, even if they are not in thermal equilibrium.

\begin{figure}
	\begin{center}
		\includegraphics[width=\columnwidth]{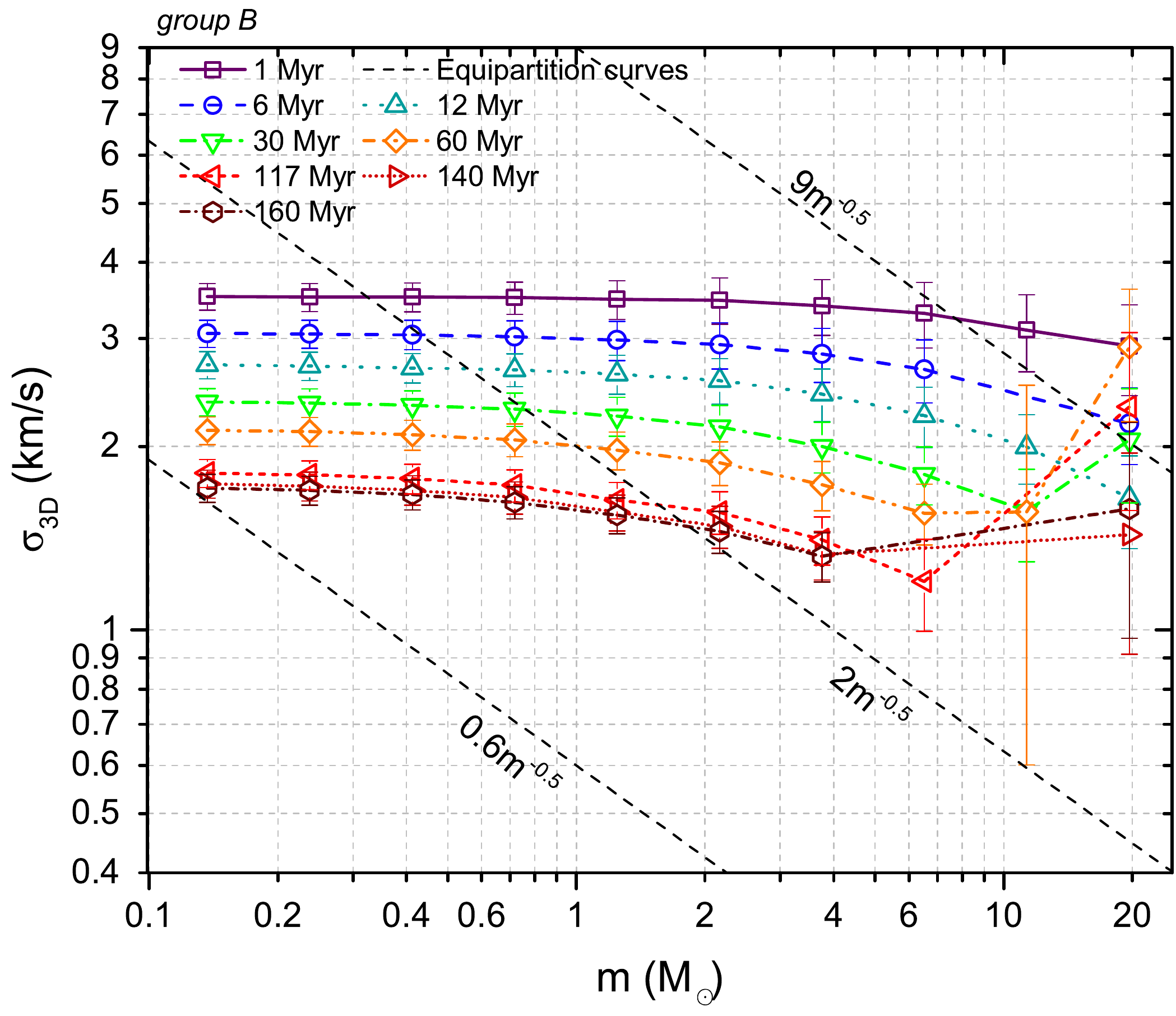}
		\caption{\label{fig:fig6} Same as Fig.~\ref{fig:fig4} but for simulations of group~B.
		}
	\end{center}
\end{figure}

Fig. \ref{fig:fig6} is the same as Fig. \ref{fig:fig4} but for the simulations of group~B, where the Galactic tidal field is not included.  From the comparison of Figs. \ref{fig:fig4} and \ref{fig:fig6} it is apparent that the Galactic tidal field  included in the simulations has no effect on the kinematic state of the stellar system. In the simulations of group~B, we estimate $\sigma_{\mathrm{3D}}\left(m\right) \propto m^{-0.07}$ for $m\lesssim 5\msun{}$ (Table~\ref{tab:tab2}). 

\begin{figure}
	\begin{center}
		\includegraphics[width=\columnwidth]{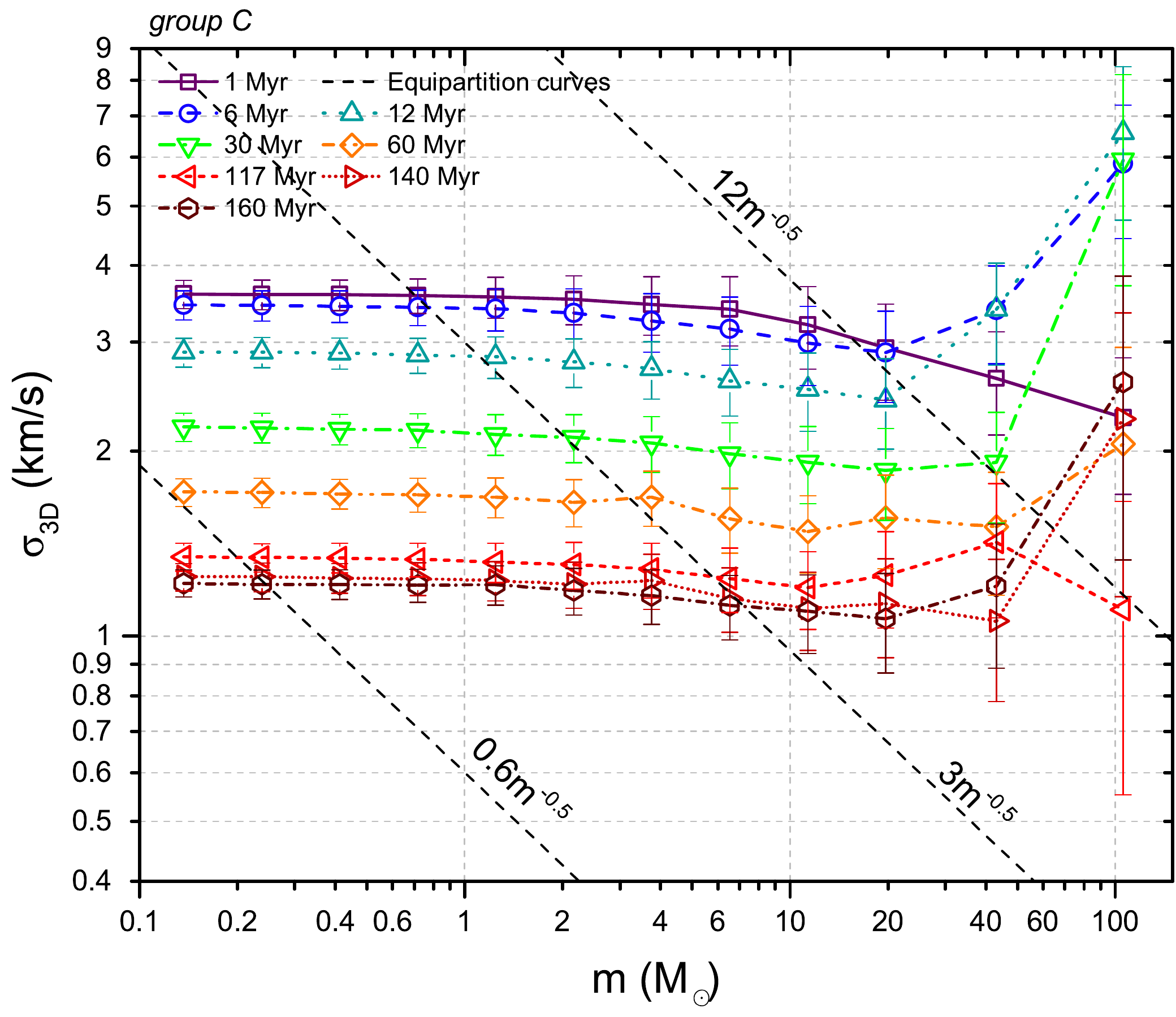}
		\caption{\label{fig:fig7} Same as Fig.~\ref{fig:fig4} but for simulations of group~C.}
	\end{center}
\end{figure}

Fig. \ref{fig:fig7} is the same as Fig. \ref{fig:fig4} but for the simulations of group~C, for which neither the Galactic tidal field nor stellar evolution are included. Without stellar evolution, the mass spectrum of the star cluster does not evolve with time. Therefore, in Fig. \ref{fig:fig7} we added two additional mass bins for stars with mass $m > 25\msun{}$. Fig. \ref{fig:fig7} is qualitatively similar to Figs. \ref{fig:fig4} and \ref{fig:fig6}: \std{} remains approximately flat for $m\lesssim 20\msun{}$ at various times. The main differences are due to the presence of very massive stars ($m>30$ M$_\odot$) for the entire simulation in runs of group~C. At time $t=1$ Myr stars more massive than $\sim{}20$ M$_\odot$ are already on their way toward equipartition. At time $t=6$ Myr the stars with mass $m>20$ M$_\odot$ have already become much hotter than lighter stars. They remain hotter for the entire simulations, with some fluctuations.

At the end of the simulation ($160$ Myr), the kinematic state of stars with mass $m\lesssim 20\msun{}$ is well described by the trend $\std{} \propto m^{-0.02}$ (Table \ref{tab:tab2}), while very massive stars have a higher velocity dispersion. We note that the velocity dispersion of the light stars at $t=160$ Myr is slightly flatter than the one found in simulations of group~A (Fig. \ref{fig:fig4}), and that the velocity dispersion of the massive stars is significantly higher.  

Moreover, at time $\gtrsim{}30$ Myr in runs C (Fig. \ref{fig:fig7}), there is no mass range close to equipartition,  in contrast with what we found in the simulations of group~A. Thus, the OCs of group~C are very far from thermal equilibrium.

\begin{figure}
	\begin{center}
		\includegraphics[width=\columnwidth]{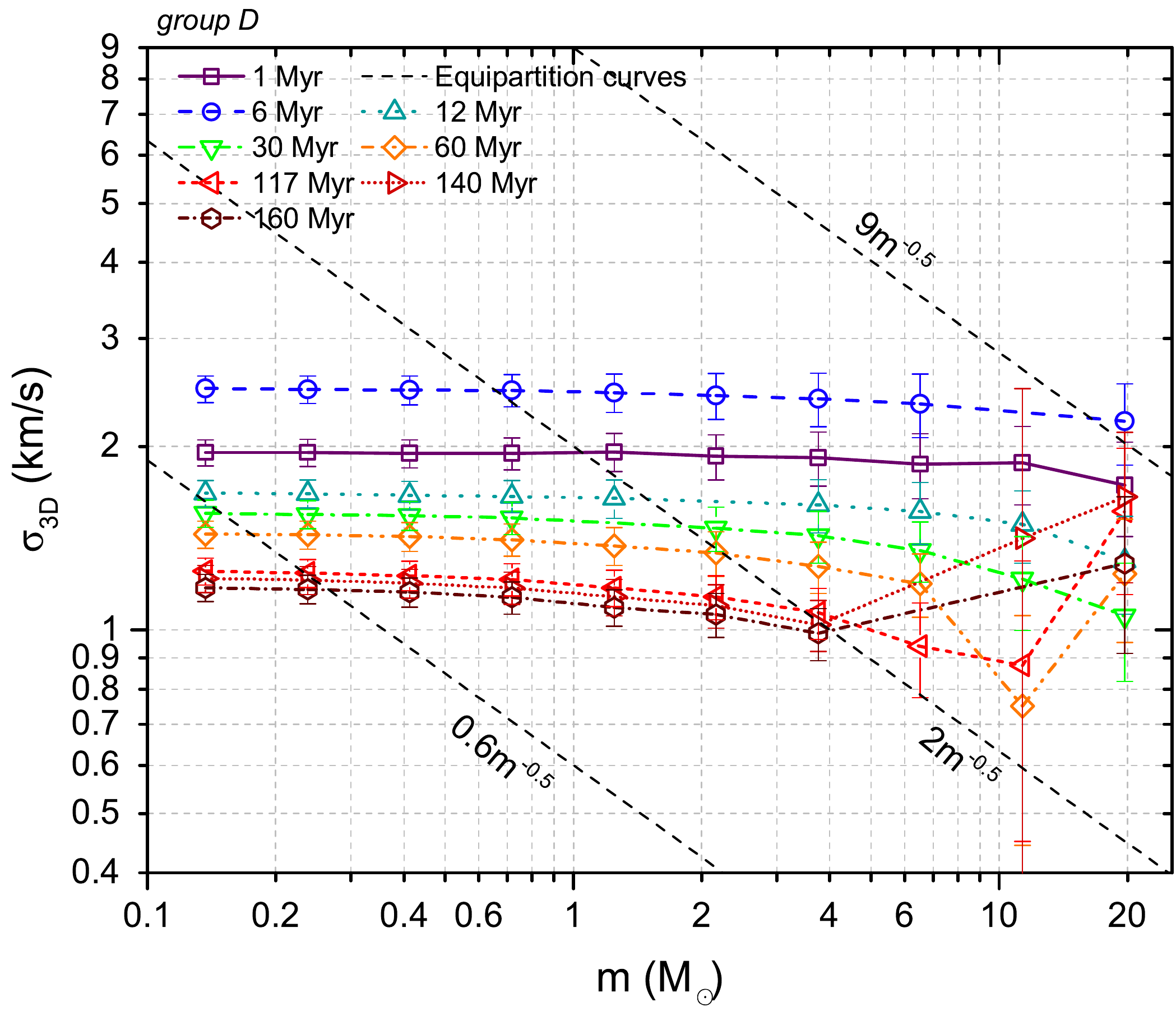}
		\caption{\label{fig:fig8} Same as Fig.~\ref{fig:fig4} but for simulations of group~D. 
		}
	\end{center}
\end{figure}

Fig. \ref{fig:fig8} is the same as Figs. \ref{fig:fig4}, \ref{fig:fig6} and \ref{fig:fig7} but for the simulations of group~D.  In this case, both stellar evolution and tidal fields are included, but each simulated star cluster is initially composed of 20 sub-clusters instead of being a monolithic system. Thus, the star cluster is not in virial equilibrium at the beginning of the simulation. Comparing Figs.  \ref{fig:fig8} and \ref{fig:fig4} we do not find significant differences in terms of kinematic state. Massive stars approximately reach thermal equilibrium at time $t\simeq 30$ Myr, later than what observed in the simulations of group~A. This happens because  the merger of the initial sub-clusters occurs during the first $\sim 15$ Myr in the simulations of group~D. Therefore, the monolithic cluster forms after $\sim 15$ Myr.  We find $\std{} \propto m^{-0.04}$ for $m\lesssim 5\msun{}$ at $\tev{}=160$ Myr (Table~\ref{tab:tab2}). As in the simulations of group~A, massive stars seem to be slightly hotter. At $t=160$ Myr, the  normalization of the \std{} curve of simulations in group~D ($\sim{}1.2$ km s$^{-1}$) is lower than that of simulations in group~A ($\sim{}1.7$ km s$^{-1}$), since the OCs of group~D have lost more mass as a consequence of the initial collapse of the 20 sub-clusters (see Table~\ref{tab:tab1a}).

We conclude that the equipartition state of a stellar system does not depend significantly on the initial spatial and velocity distribution of stars, at least for the initial conditions investigated in this paper.

\begin{figure}
	\begin{center}
		\includegraphics[width=\columnwidth]{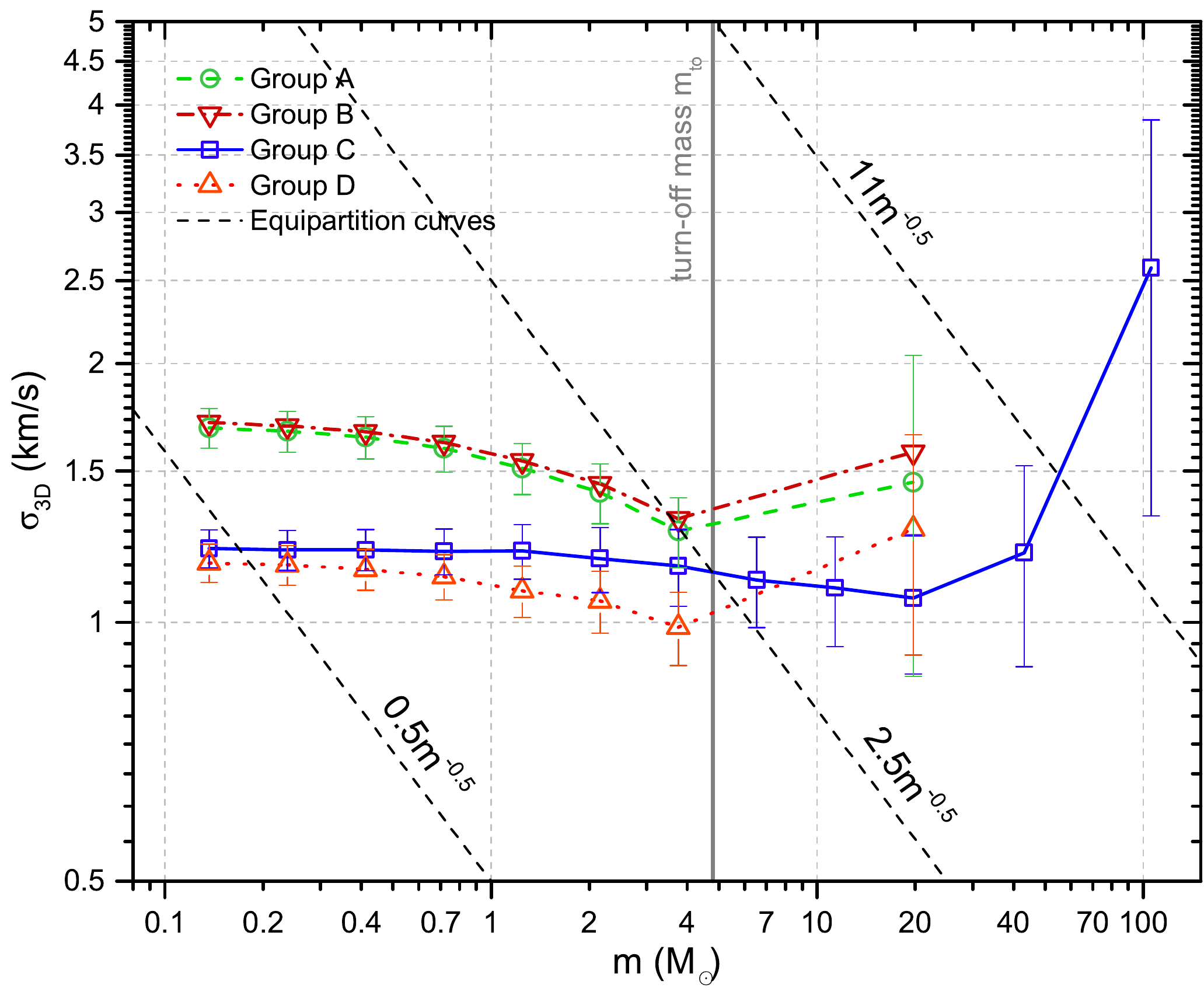}
		\caption{\label{fig:fig9} Three-dimensional velocity dispersion \std{}, as a function of mass, for the simulations of all the groups, at time $t= 160$ Myr $\simeq 6 t_{\mathrm{rh}}\left(0\right)$. The dashed black lines are the family of equipartition curves $\sigma\left(\alpha;m\right) = \alpha m^{-0.5}$ and the solid gray line,  at $m = m_{\mathrm{to}} = 4.8\msun{}$, indicates the turnoff mass at $t = 160$ Myr. The dashed green line with circles refers to the simulations of group~A. Dash-dot red curve with downward triangles: group~B. Solid blue line with squares: group~C. Dotted orange curve with upward triangles: group~D.}
	\end{center}
\end{figure}

Fig. \ref{fig:fig9} shows the velocity dispersion as a function of mass for all simulation groups, at time $t=160$ Myr $\simeq 6\,{} t_{\mathrm{rh}}\left(0\right)$. Tab. \ref{tab:tab2} lists the best fit coefficients of the function $\std{} = c\,{}m^{\beta}$ for the \std{} trends shown in Fig. \ref{fig:fig9}, for $m\lesssim \mto{}$. 

Fig. \ref{fig:fig9} and Tab. \ref{tab:tab2} help us summarizing the main results:

\begin{itemize}
	\item[-] the simulated star clusters show a clear signature of mass segregation (see Fig. \ref{fig:fig1});
	\item[-] after several initial half-mass relaxation times, all simulated star clusters are significantly far from kinetic energy equipartition (see Fig. \ref{fig:fig9}); 
         \item[-] the velocity dispersion does not seem to depend on the mass for most of the considered mass spectrum and for all considered simulation groups; only very massive objects ($\gtrsim{}20$ M$_\odot$) seem to be dynamically hotter than lighter stars (see Fig. \ref{fig:fig9}); 
	\item[-] thermal equilibrium is not reached either locally (i.e. in a radial annulus of the star clusters) or globally (see Fig. \ref{fig:fig5});
	\item[-] starting from clumpy instead of monolithic initial conditions has a mild effect on the final kinematic state of the star cluster;
	\item[-] the presence of a Galactic tidal field has no effect on the final kinematic state of the star cluster, at least for a moderate mass ($\sim{}4000$ M$_\odot$) OC in the solar neighborhood;
	\item[-] in the absence of stellar evolution, the final \std{} curve of stars with mass $m\lesssim 20\msun{}$ tends to be even flatter and very massive stars are significantly hotter than what we found in runs with stellar evolution;
        \item[-] the normalization of \std{} for stars with mass $m\lesssim 5\msun{}$ in the simulations of groups~C and D is a factor of $\sim{}1.3$ lower than that of groups~A and B. For the OCs of group~D the reason is that the initial violent merger of the 20 sub-clusters leads to the ejection of more stars from the stellar systems (Tab.~\ref{tab:tab1a}). In contrast, the OCs of group~C have approximately the same final mass as those of group~A and B but they have a $\sim{}1.8$ times larger final half-mass radius (Tab.~\ref{tab:tab1a}), which corresponds to a factor of $\sim{}1.3$ difference in the velocity dispersion (for OCs in virial equilibrium).
\end{itemize} 

\begin{table} 
	\begin{center}
		\caption{\label{tab:tab2} Coefficients $c$ and $\beta$ of the fits $\std{} = cm^{\beta}$ and corresponding errors $c_{\mathrm{err}}$ and $\beta_{\mathrm{err}}$, for all simulation groups, at $t=\tev{}$, for $m\lesssim4.8\msun{}=\mto{}$.}
		\begin{tabular}{ l l l l l }
			\hline
			\myalign{c}{Group} & \myalign{c}{$c$} & \myalign{c}{$c_{\mathrm{err}}$} & \myalign{c}{$\beta$} & \myalign{c}{$\beta_{\mathrm{err}}$} \\ 
			\hline
			
			\myalign{c}{A} & $1.505$ & $0.024$ & $-0.071$ & $0.012$ \\ 
			\myalign{c}{B} & $1.536$ & $0.022$ & $-0.067$ & $0.011$  \\ 
			\myalign{c}{C} & $1.1944$ & $0.0063$ & $-0.0138$ & $0.0040$  \\ 
			\myalign{c}{D} & $1.094$ & $0.010$ & $-0.0435$ & $0.0073$\\
			\hline
		\end{tabular}
	\end{center}
\end{table}


\section{Discussion}
\label{sec:discussion}
In this section we discuss why the simulated OCs do not attain kinetic energy equipartition. Five physical processes play an important role in the interpretation of our results:

\begin{itemize}
	\item[-] the Spitzer's instability, applied to a stellar system with a realistic mass function;
        \item[-] mass segregation;
        \item[-] core collapse;
	\item[-] the formation and dynamical evolution of binary systems;
	\item[-] mass loss by stellar winds.
\end{itemize}

At the beginning of the simulations, the massive stars interact with the light stars,  their velocity dispersion decreases and they sink toward the centre of the stellar system via dynamical friction. Thus, the heaviest stars are the first ones that tend to equipartition. For the simulations of group~A, this process goes on until stars with mass $m\gtrsim 10\msun{}$ reach equipartition (see curves between $t \simeq 1 $ Myr and $t\simeq 12$ Myr of Fig. \ref{fig:fig4}).

\begin{figure}
	\begin{center}
		\includegraphics[width=\columnwidth]{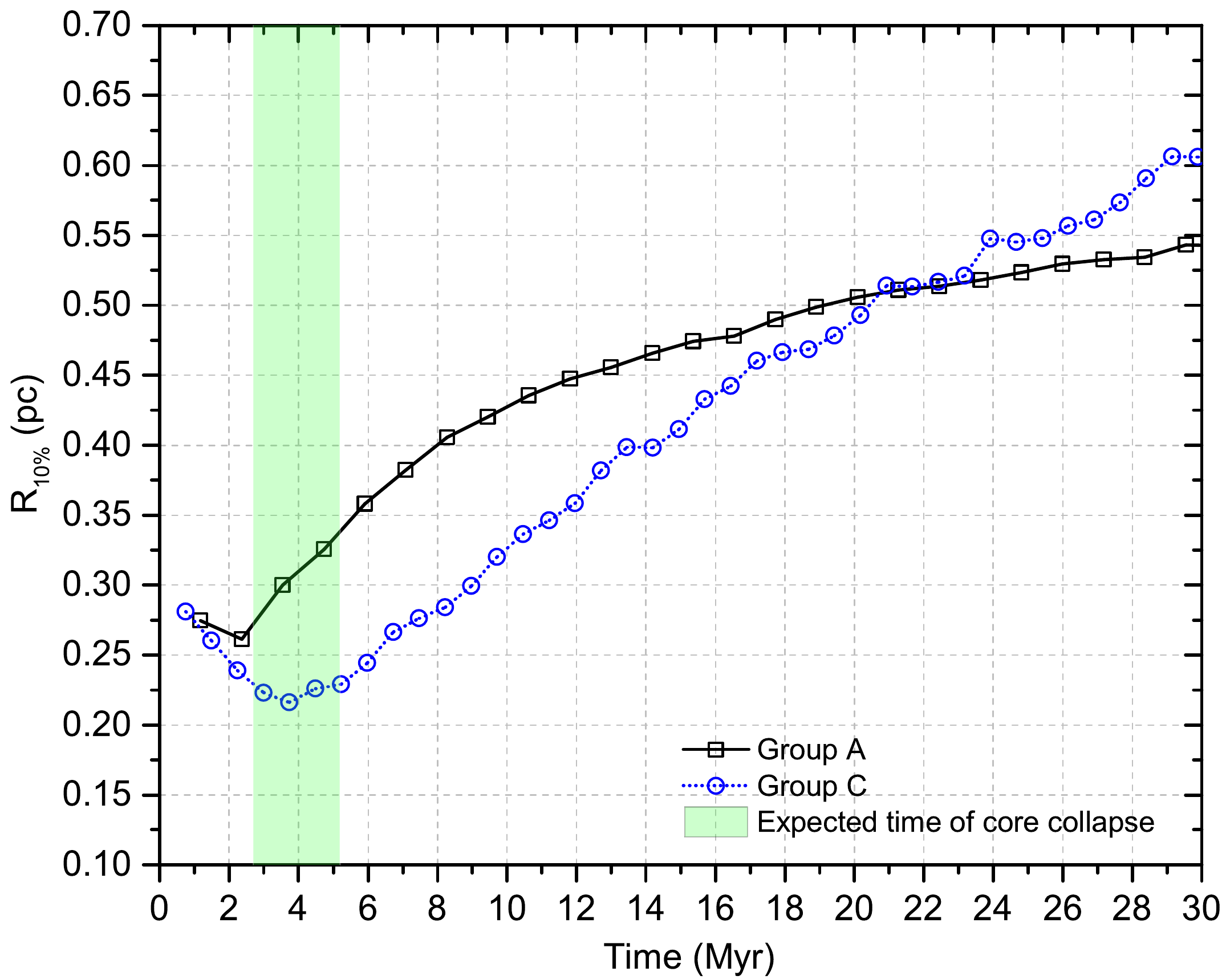}
		\caption{\label{fig:fig10} Lagrangian radius that contains $10$\% of the total mass of the star cluster, as a function of time. The solid black line with squares refers to the simulations of group~A while the dotted blue line with circles refers to the simulations of group~C. Each line is averaged over all simulations of the same group. The green semi-transparent area highlights the interval of time in which we expect to observe core collapse for the star clusters of group~C.}
	\end{center}
\end{figure}

A few Myr after the beginning of the simulation, the star cluster becomes mass segregated, evolves toward core collapse (Fig.~\ref{fig:fig10}), and the gravitational interactions between stars, especially in the inner regions, become more and more frequent. As a consequence, the velocity dispersion of the massive stars starts increasing while that of lighter stars is unchanged. In particular, the \std{} of massive objects rises up  at $\sim 30-60$ Myr in simulations of group~A (Fig.~\ref{fig:fig4}) and B (Fig.~\ref{fig:fig6}), at $\sim 6$ Myr in those of group~C (Fig.~\ref{fig:fig7}) and  at $\sim 60-120$ Myr in those of group~D (Fig.~\ref{fig:fig8}). The rise of \std{} occurs much earlier in group~C than in the other simulations, because stellar winds are switched off. In fact, when stellar winds are effective, they make the cluster potential well shallower and drive an expansion of the core, reducing the efficiency of close dynamical encounters (see \citealt{trani2014} and \citealt{mapelli2016} for details). Moreover, mass segregation is faster in runs of group~C, because very massive stars do not lose mass in the first Myr and efficiently sink to the centre by dynamical friction.

The rise of \std{}  is not due to BH natal kicks. In fact, \std{} increases more in runs of group~C (Fig. \ref{fig:fig7}), where stellar evolution and natal kicks are switched off, than in the other runs. Moreover, all BHs with mass $\gtrsim{} 13\,{}\msun{}$ do not receive kicks, since they form through direct collapse \citep{spera2015}. Still, BHs with mass $\gtrsim{} 13\,{}\msun{}$ show a velocity dispersion higher than lighter stars. The effect of natal kicks is to produce a dearth of compact remnants with mass between $6\msun{}$ and $13\,{}\msun{}$, because they are ejected from $r_{\rm h}$. In addition, the ejection of light ($<13\,{}\msun$) compact remnants from the core by natal kicks contributes to the expansion of the core, and thus has the same effect as stellar winds: it delays the rise of \std{} of very massive objects (see Fig.~6 of \citealt{mapelli2013b}).

The core collapse is reversed by mass loss due to stellar winds and SNe \citep{mapelli2013,trani2014}, when these are included in the simulations, and by three-body encounters with a massive binary. Since our simulations do not include primordial binaries, binaries form dynamically when the central density increases.  

\begin{figure}
	\begin{center}
		\includegraphics[width=\columnwidth]{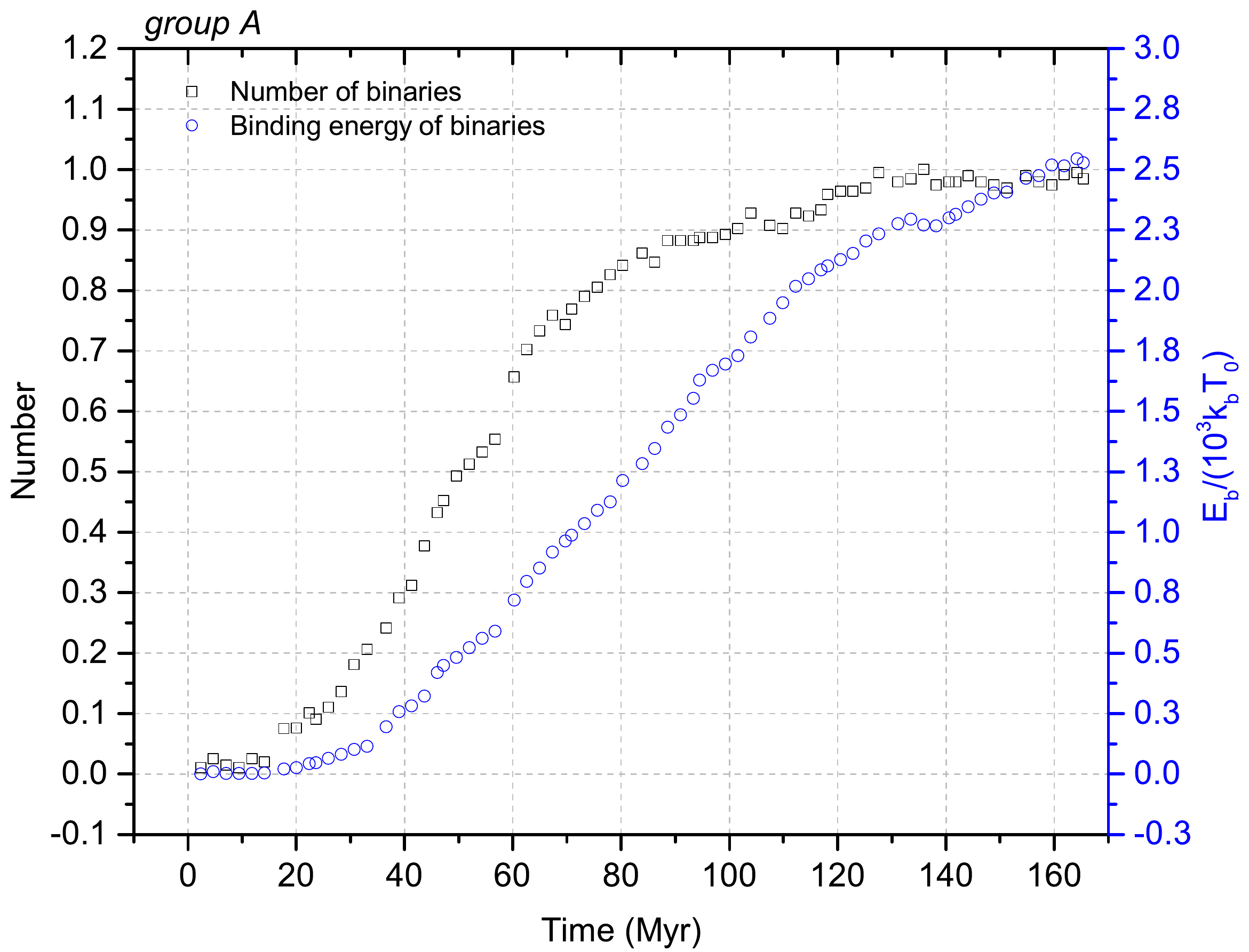}
		\caption{\label{fig:fig11} Average number of binaries per star cluster (left y-axis, black open squares) and average binding energy per binary normalized to $10^3k_B\,{}T\left(t=0\right)$ (right x-axis, blue open circles), as a function of time, for the simulations of group~A.}
	\end{center}
\end{figure}

\begin{figure}
	\begin{center}
		\includegraphics[width=\columnwidth]{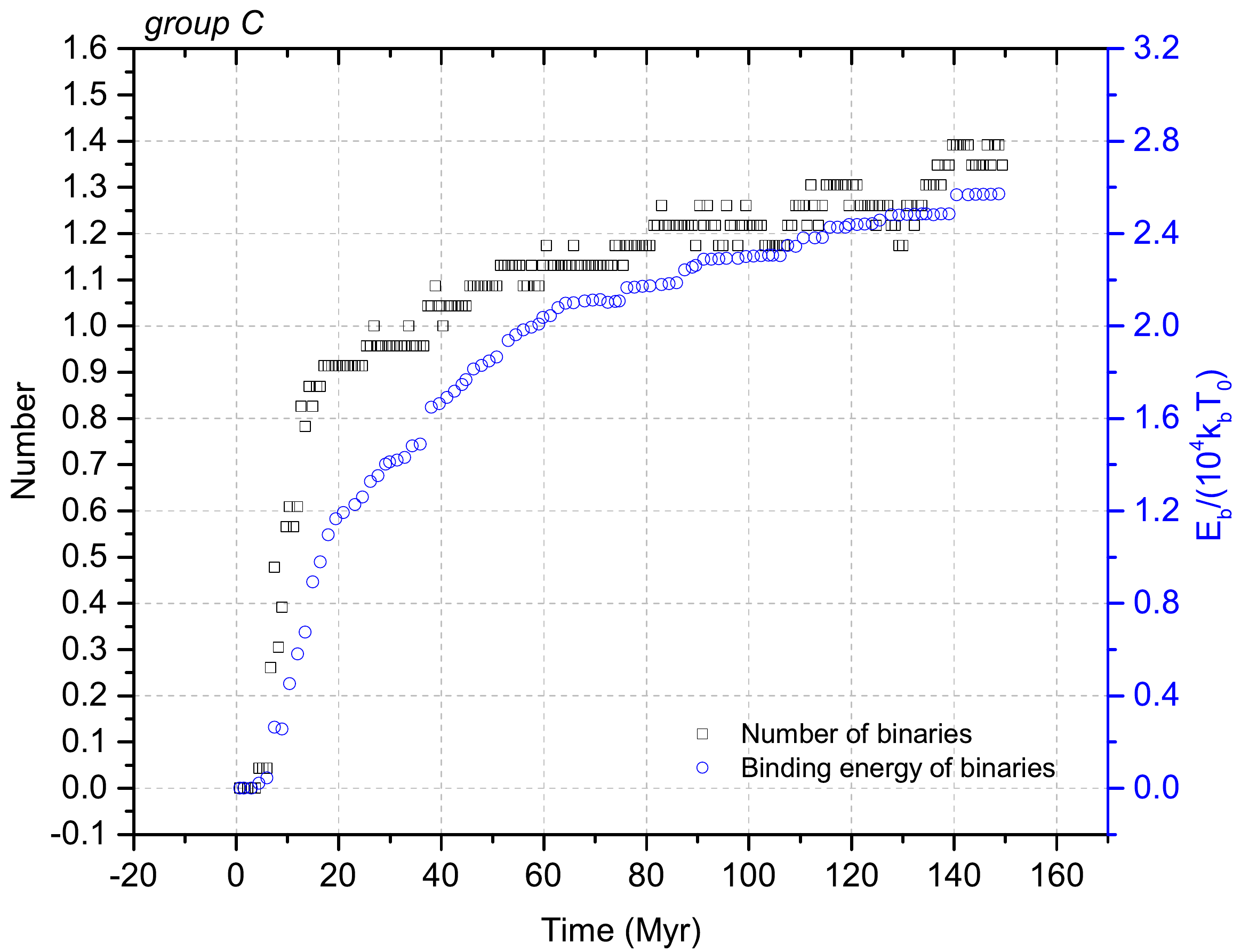}
		\caption{\label{fig:fig12} Same as Fig. \ref{fig:fig11} but for the simulations of group~C. The normalization of binding energy here is $10^4k_B\,{}T\left(t=0\right)$.}
	\end{center}
\end{figure}

Figs.~\ref{fig:fig11} and  ~\ref{fig:fig12} show the average number of binaries and their average energy as a function of time, for groups A and C, respectively. More than one binary might form dynamically, but in most simulations there is only $\sim{}1$ hard binary in the core at a given time (a binary is hard if its binding energy is larger than the average kinetic energy of a star in the OC, \citealt{heggie1975}). When a hard binary merges, breaks by SN explosion, or is kicked out off the OC by dynamical interactions, a new hard binary forms in the core and replaces the previous one. A hard binary exchanges energy with the passing-by stars through three-body interactions and acts as energy reservoir,  keeping the system stable against the gravothermal catastrophe.

Simulations of group~C undergo a strong core collapse between $\sim{}2$ and $\sim{}6$ Myr  (Fig.~\ref{fig:fig10}). In runs of group~C, the formation of the first hard binary  occurs on a very short timescale ($t\sim{}5-15$ Myr, Fig.~\ref{fig:fig12}) and coincides with the core collapse. In fact,  there is no stellar evolution in simulations of group~C. In absence of stellar winds, three-body encounters are the only mechanism able to reverse the core collapse. 

In contrast, simulations of group~A  undergo only a mild core collapse at $\sim{}2$ Myr, that is immediately reversed. As already showed by \citet{mapelli2013} and \citet{trani2014}, the first core collapse is reversed by stellar winds, even without the formation of binaries. Fig. \ref{fig:fig11} shows that, for $t\lesssim 15$ Myr, more than 98\% of OCs in group~A do not contain any binary systems. At $\gtrsim{}10$ Myr the stellar winds by massive stars  are over, and the core tends to collapse again. At this stage, hard binaries start forming, and three-body encounters keep the core stable against further collapses. Simulations of group~B and D evolve in a similar way as simulations of group~A.

 The time when the first binary forms (from Figs.~\ref{fig:fig11} and  ~\ref{fig:fig12} for A and C, respectively) is remarkably similar to the time when the \std{} of the most massive stars upturns, moving away from equipartition (from Figs.~\ref{fig:fig4} and  ~\ref{fig:fig7} for A and C, respectively).

For the simulations of group~A, the first binary forms  between $t\simeq 10$ Myr and $t\simeq 80$ Myr. At this stage,  the simulated OCs are mass segregated, therefore the hard binary transfers kinetic energy mainly to the surrounding massive stars. Most of the light stars reside in the outer parts of the OCs and do not undergo gravitational encounters. This means that the heaviest particles interact with each other, ignoring the  light stars.  Three-body gravitational scatters with the hard binaries increase the velocity dispersion of massive stars. This drives Spitzer's instability, and explains the fact that the \std{} curve of massive stars ($m\gtrsim 12\msun{}$) rises up between 12 Myr and 60 Myr  in the simulations of group~A (see Fig. \ref{fig:fig4}). 

The hard binary transfers kinetic energy to the surrounding stars, hardens and increases its binding energy. The kinetic energy released by the central binary system let the cluster expand, therefore the frequency of three-body encounters and the need to release further kinetic energy decrease. In fact, Fig. \ref{fig:fig11} shows that the average number of binaries becomes nearly constant  at $t\simeq 110$ Myr. For $t \gtrsim 110$ Myr the hard binary system keeps transferring energy to passing-by stars and keeps shrinking (the binding energy keeps increasing) and no more binaries form. Since there are only few light stars in the central region, kinetic energy is transferred mainly to massive stars, that cannot reach equipartition anymore. On the other hand, light stars ignore the complex evolution that happens inside the core, so their kinematic state is approximately unchanged. This mechanism explains why the \std{} trend for light stars seems to be almost unaffected since the formation of the first binary system and why massive stars seem to have a higher velocity dispersion.

The  process we described above occurs in all simulation groups. In fact, the \std{} plots showed in Sec. \ref{sec:results} for the different simulation groups are qualitatively similar to each other at $t=6$ $t_{\rm rh}$. Simulations of group~C are the ones that show the main differences with respect to the other runs (i.e. group~A, B and D). In the simulations of group~C (where the stellar evolution is turned off)  Spitzer's instability is stronger and develops earlier than in the other groups. This happens because the mass of very massive stars remains constant for the entire simulations, instead of decreasing by stellar winds. This enhances the efficiency of mass segregation and leads to a fast core collapse. 

A tight binary system forms significantly earlier than in the other simulations, because stellar winds do not contribute to reverse the core collapse.  Moreover, there are no SN kicks that eject massive objects from the core. Thus, the evolution of the core is completely dominated by very massive stars that interact with each other for the entire simulation. Their velocity dispersion grows because of these interactions. The light stars lie mostly in the outskirts of the star clusters, and are not affected by what happens in the central regions.
	
In this section, we highlighted the importance of three-body encounters with binary systems. However, our simulations do not include primordial binaries. This might seem a severe problem of our model, because the binary fraction in OCs is very high \citep{sollima2010,li2013}. However, \citet{mapelli2013} showed that the overall kinematic evolution of an OC is not affected by the number of primordial binaries (see e.g. their figure 8). The main reason is that a primordial binary does not transfer significant energy to the other stars of the cluster unless the cluster evolves toward gravothermal instability. During the gravothermal instability, hard binaries transfer enough kinetic energy to reverse the collapse of the core, and to keep it stable. The energy needed to reverse the core collapse depends on the structural properties of the cluster and does not depend on the number of binaries  (e.g. \citealt{goodman1987}). 

If there are primordial binaries, they start transferring kinetic energy during the core collapse (not before); if there are no primordial binaries, enough binaries form dynamically to sustain the core against collapse. Thus, the energy that is exchanged between stars does not depend on the number of binaries. This means that our main conclusions about equipartition are not affected by the binary fraction.

\section{Summary}
\label{sec:summary}

In this paper, we investigate energy equipartition in OCs by means of direct-summation N-body simulations. {We adopt a \citet{kroupa2001} initial mass function. We start from both virial and sub-virial initial conditions, and we check the effects of the Galactic tidal field and of stellar evolution.

We find that energy equipartition is not attained by the simulated stellar systems, even if we integrate them for $\sim{}6$ two-body relaxation timescales (Fig. \ref{fig:fig9}).  Moreover, energy equipartition is reached  neither locally (i.e. in a radial annulus of the star clusters) nor globally (see Fig. \ref{fig:fig5}). 

For most of the  mass spectrum, the velocity dispersion $\sigma{}(m)$ does not depend  on the star mass in all our simulations. Only very massive objects ($\gtrsim{}20$ M$_\odot$) seem to be dynamically hotter than lighter stars. This is in contrast  with the equipartition theory, which predicts that $\sigma{}(m)\propto{}m^{-0.5}$. 

This result is not affected by the Galactic tidal field, at least for a moderately massive OC ($\sim{}4000$ M$_\odot$) in the solar neighborhood. Moreover, this result does not depend on the initial spatial distribution function. In fact, we obtain the same trend of the velocity dispersion \std{} if we start from a monolithic King model or from clumpy sub-virial initial conditions. 

In simulations without stellar evolution, the final \std{} curve of stars with mass $m\lesssim 20\msun{}$ tends to be even flatter and very massive stars are significantly  hotter than what we found in runs with stellar evolution  (Fig. \ref{fig:fig9}).

All simulated star clusters are significantly mass segregated by the end of the simulations (Fig. \ref{fig:fig1}). Thus, dynamical friction is efficient, even if energy equipartition is not achieved.

In the first stages of their evolution ($<t_{\rm rh}$), the simulated OCs seem to evolve toward equipartition. In fact, the most massive stars interact with the other stars, transferring kinetic energy to the lighter ones, and sink toward the centre by dynamical friction. 

When the OC becomes significantly mass segregated, its core develops gravothermal instability. This increases the central stellar density and leads to the formation of hard binaries, which transfer kinetic energy to the other stars and keep the core stable against further collapse. Stellar winds, when included in the simulations, contribute to reverse the first core collapse and delay the formation of hard binaries. 

The time when hard binaries form to prevent core collapse ($t<15$ Myr in simulations without stellar evolution, $t>15$ Myr in the other simulations) coincides with the time when the OC stops evolving toward equipartition. At this stage, the velocity dispersion of the most massive stars becomes higher than the velocity dispersion of the lighter stars. 

We interpret this fact as a consequence of the strong mass segregation in the centre of the OC: when the hard binaries form and reverse the gravothermal instability by transferring kinetic energy to the surrounding stars via three-body encounters, the core is populated mainly by the most massive stars, whereas the lighter stars are in the OC outskirts. Thus, the hard binaries transfer kinetic energy mostly to massive stars and stellar remnants, whose velocity dispersion increases. This effect is reminiscent of Spitzer's instability, but for a realistic initial mass function.

The fact that the velocity dispersion of massive stars is particularly high in the runs without stellar winds confirms our interpretation. In fact, stellar winds remove mass from the core of the cluster, making its potential well shallower and preventing core collapse \citep{mapelli2013,trani2014}. Moreover, stellar winds slow down the process of mass segregation, because the most massive stars lose most of their mass in a few Myr. In absence of stellar winds, both mass segregation and core collapse are particularly strong. To reverse core collapse, the most massive objects need to acquire more kinetic energy by close encounters, and become hotter than they do in the simulations with stellar winds.

We note that Spitzer's instability is important especially for BHs. This has crucial implications for the detection of gravitational waves by the Advanced  LIGO and Virgo detectors. In a follow-up study, we will quantify the impact of our result on the detection  of gravitational waves by Advanced  LIGO and Virgo.

Our simulations strongly support the result that OCs do not attain equipartition, regardless of their initial conditions. Data from the GES and from the Gaia mission are essential to confirm this result.

\section{Acknowledgments}
 We thank the anonymous referee for suggestions that helped us to
improve the paper. MM and MS thank Alessandro Alberto Trani, Paolo Miocchi, Germano Sacco and Alessandro Bressan for useful discussions. We made use of the \starlab{} (ver. 4.4.4) public software environment and of the \sapporo{} library. We thank the developers of  \starlab{}, and especially its primary authors (P. Hut, S. McMillan, J. Makino, and S. P. Zwart). We thank the authors of \sapporo{}, and in particular E. Gaburov, S. Harfst, and S. Portegies Zwart. MM and MS acknowledge financial support from the Italian Ministry of Education, University and Research (MIUR) through grant FIRB 2012 RBFR12PM1F and from INAF through grant PRIN-2014-14. MM acknowledges financial support from the MERAC Foundation.


\appendix
\section{Late evolution of OCs}

	
	\begin{figure}
		\begin{center}
			\includegraphics[width=\columnwidth]{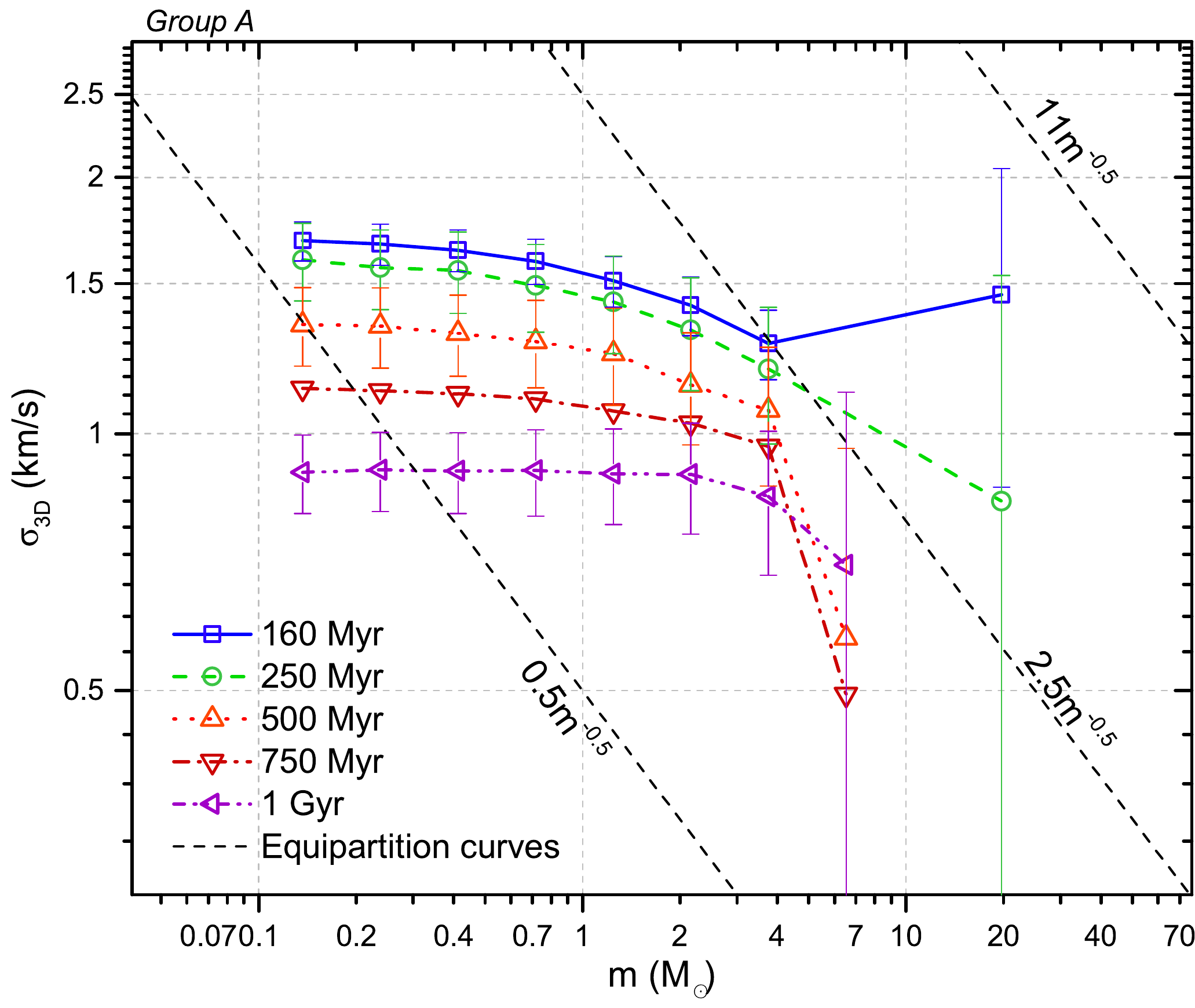}
			\caption{\label{fig:fig3b} Three-dimensional velocity dispersion \std{}, as a function of mass, for the simulations of group~A, at time $t=160$ Myr (solid blue line with open squares), $t=250$ Myr (dashed green line with open circles), $t=500$ Myr (dotted orange line with upward triangles), $t=750$ Myr (dash-dot red line with downward triangles), and $t=1$ Gyr (dash-dot-dot purple line with leftward triangles). The dashed black lines are the family of equipartition curves $\sigma\left(\alpha;m\right) = \alpha \,{} m^{-0.5}$.}
		\end{center}
	\end{figure}
 We simulated 20 out of the 200 runs of group~A for 1 Gyr $\simeq 40 t_{\mathrm{rh}}\left(0\right)$ (rather than 160 Myr), to check whether equipartition can be reached at later times.	
Fig. \ref{fig:fig3b} shows the velocity dispersion curves as a function of mass for these simulations at several times. Even at $t=1$ Gyr, the OCs are far from kinetic energy equipartition. The \std{} curve is still flat for stars with $m \lesssim 3\msun{}$ and heavy stars with mass between $4\msun{}$ and $8\msun{}$ are slightly closer to equipartition. It is worth noting that the last point in the \std{} curves fluctuates significantly since we do not have enough statistics for these runs.

Moreover, there is not enough statistics to plot a data point at $m\gtrsim{}10\msun{}$. The reason is that most BHs have been ejected out of the half-mass radius at times $t\gtrsim 500$ Myr. This is a further confirmation that Spitzer instability is effective in our OCs (see e.g. \citealt{sigurdsson1993}). In a follow-up study, we will investigate the ejection history of dark remnants in our simulations.

\bibliography{equipartition}

\end{document}